\documentclass[onecolumn,prd,nofootinbib,aps,floats,floatfix,amsmath,amssymb,secnumarabic,preprintnumbers,showpacs,notitlepage]{revtex4-1} %
\usepackage[final]{graphicx}

\def\beq{\begin{equation}}
\def\eeq{\end{equation}}
\def\bea{\begin{eqnarray}}
\def\eea{\end{eqnarray}}
\def\nn{\nonumber}
\def\roughly#1{\mathrel{\raise.3ex\hbox{$#1$\kern-.75em\lower1ex\hbox{$\sim$}}}}
\def\lsim{\roughly<}
\def\gsim{\roughly>}

\def\bs{B^0_s}
\def\bsbar{{\bar B}^0_s}
\def\bsmumu{b \to s \mu^+ \mu^-}
\def\bsee{b \to s e^+ e^-}
\def\bsnunubar{b \to s \nu {\bar\nu}}

\def\BKstarmumu{B \to K^* \mu^+ \mu^-}

\def \SM{{\rm SM}}

\def \expt{{\rm expt}}

\def\s{\sqrt{2}}
\def\bsll{b \to s \ell^+ \ell^-}

\newcommand{\av}[1]{\langle #1 \rangle}

\def\gsim{{~\raise.15em\hbox{$>$}\kern-.85em
          \lower.35em\hbox{$\sim$}~}}
\def\lsim{{~\raise.15em\hbox{$<$}\kern-.85em
          \lower.35em\hbox{$\sim$}~}}

\newcommand{\gev}{\ensuremath{\mathrm{\,Ge\kern -0.1em V}}}

\def\Z{Z^{\prime}}

\begin{document}

\title{\boldmath New Physics in $\bsmumu$ after the Measurement of $R_{K^*}$}
\preprint{UdeM-GPP-TH-17-256; WSU-HEP-1708}
\author{Ashutosh Kumar Alok}
\email{akalok@iitj.ac.in}
\affiliation{Indian Institute of Technology Jodhpur, Jodhpur 342011, India}
\author{Bhubanjyoti Bhattacharya}
\email{bbhattach@ltu.edu}
\affiliation{Department of Natural Sciences, Lawrence Technological University, Southfield, MI 48075, USA and 
Department of Physics and Astronomy, Wayne State University, Detroit, MI 48201, USA}
\author{Alakabha Datta}
\email{datta@phy.olemiss.edu}
\affiliation{Department of Physics and Astronomy, 108 Lewis Hall, \\
University of Mississippi, Oxford, MS 38677-1848, USA}
\author{Dinesh Kumar}\email{dinesh@phy.iitb.ac.in}
\affiliation{Indian Institute of Technology Bombay, Mumbai 400076, India}
\affiliation{Department of Physics, University of Rajasthan, Jaipur 302004, India}
\author{Jacky Kumar}\email{jka@tifr.res.in}
\affiliation{Department of High Energy Physics, Tata Institute of Fundamental Research, \\
400 005, Mumbai, India}
\author{David London}\email{london@lps.umontreal.ca}
\affiliation{Physique des Particules, Universit\'e de Montr\'eal, \\
C.P. 6128, succ. centre-ville, Montr\'eal, QC, Canada H3C 3J7}


\begin{abstract}
The recent measurement of $R_{K^*}$ is yet another hint of new physics
(NP), and supports the idea that it is present in $\bsmumu$ decays. We
perform a combined model-independent and model-dependent analysis in
order to deduce properties of this NP. Like others, we find that the
NP must obey one of two scenarios: (I) $C_9^{\mu\mu}({\rm NP}) < 0$ or
(II) $C_9^{\mu\mu}({\rm NP}) = - C_{10}^{\mu\mu}({\rm NP}) < 0$. A
third scenario, (III) $C_9^{\mu\mu}({\rm NP}) = - C_{9}^{\prime
  \mu\mu}({\rm NP})$, is rejected largely because it predicts $R_K =
1$, in disagreement with experiment. The simplest NP models involve
the tree-level exchange of a leptoquark (LQ) or a $Z'$ boson. We show
that scenario (II) can arise in LQ or $Z'$ models, but scenario (I) is
only possible with a $Z'$. Fits to $Z'$ models must take into account
the additional constraints from $\bs$-$\bsbar$ mixing and neutrino
trident production. Although the LQs must be heavy, O(TeV), we find
that the $Z'$ can be light, e.g., $M_{Z'} = 10$ GeV or 200 MeV.
\end{abstract}



\maketitle

\section{Introduction}
\label{Sec:Intro}

The LHCb Collaboration recently announced that it had measured the
ratio $R_{K^*} \equiv {\cal B}(B^0 \to K^{*0} \mu^+ \mu^-)/{\cal
  B}(B^0 \to K^{*0} e^+ e^-)$ in two different ranges of the dilepton
invariant mass-squared $q^2$ \cite{RK*expt}. The result was
\beq
R_{K^*}^\expt =
\left\{
\begin{array}{cc}
0.660^{+0.110}_{-0.070}~{\rm (stat)} \pm 0.024~{\rm (syst)} ~,~~ & 0.045 \le q^2 \le 1.1 ~{\rm GeV}^2 ~, \\
0.685^{+0.113}_{-0.069}~{\rm (stat)} \pm 0.047~{\rm (syst)} ~,~~ & 1.1 \le q^2 \le 6.0 ~{\rm GeV}^2 ~.
\end{array}
\right.
\label{RK*meas}
\eeq
In the SM calculation of $R_{K^*}$ \cite{RK*theory}, the
  effect of the mass difference between muons and electrons is
  non-negligible only at very small $q^2$. As a consequence, the SM
  predicts $R_{K^*}^\SM \simeq 0.93$ at low $q^2$ \cite{flavio}, but
  $R_{K^*}^\SM \simeq 1$ elsewhere. The measurements then differ from
  the SM prediction by 2.2-2.4$\sigma$ (low $q^2$) or 2.4-2.5$\sigma$
(medium $q^2$), and are thus hints of lepton flavor
non-universality. These results are similar to that of the LHCb
measurement of $R_K \equiv {\cal B}(B^+ \to K^+ \mu^+ \mu^-)/{\cal
  B}(B^+ \to K^+ e^+ e^-)$ \cite{RKexpt}:
\beq
R_K^\expt = 0.745^{+0.090}_{-0.074}~{\rm (stat)} \pm 0.036~{\rm (syst)} ~,~~ 1 \le q^2 \le 6.0 ~{\rm GeV}^2 ~,
\label{RKexpt}
\eeq
which differs from the SM prediction of $R_K^\SM = 1 \pm 0.01$
\cite{IsidoriRK} by $2.6\sigma$.

If new physics (NP) is indeed present, it can be in $\bsmumu$ and/or
$\bsee$ transitions. In the case of $R_K$, the measurement of ${\cal
  B}(B^+ \to K^+ e^+ e^-)$ was found to be consistent with the
prediction of the SM, suggesting that the NP is more likely to be in
$\bsmumu$. However, for $R_{K^*}$, based on the information given in
Ref.~\cite{RK*expt}, a similar conclusion cannot be drawn. In any
case, it must be stressed that there are important theoretical
uncertainties in the SM predictions for ${\cal B}(B \to K^{(*)} \ell^+
\ell^-)$ ($\ell = e,\mu$) \cite{bslltheorerror}, so it is difficult to
identify experimentally whether $\bsmumu$ or $\bsee$ has been affected
by NP. On the other hand, the theoretical uncertainties essentially
cancel in both $R_{K^*}$ and $R_K$, making them very clean probes of
NP.

There are several other measurements of $B$ decays that are in
disagreement with the predictions of the SM, and these involve only
$\bsmumu$ transitions:
\begin{enumerate}

\item $B \to K^* \mu^+\mu^-$: The LHCb
  \cite{BK*mumuLHCb1,BK*mumuLHCb2} and Belle \cite{BK*mumuBelle}
  Collaborations have made measurements of $B \to K^* \mu^+\mu^-$.
  They find results that deviate from the SM predictions, particularly
  in the angular observable $P'_5$ \cite{P'5}. Recently, the ATLAS
  \cite{BK*mumuATLAS} and CMS \cite{BK*mumuCMS} Collaborations
  presented the results of their measurements of the $B \to K^*
  \mu^+\mu^-$ angular distribution.

\item $\bs \to \phi \mu^+ \mu^-$: LHCb has measured the branching
  fraction and performed an angular analysis of $\bs \to \phi \mu^+
  \mu^-$ \cite{BsphimumuLHCb1,BsphimumuLHCb2}. They find a $3.5\sigma$
  disagreement with the predictions of the SM, which are based on
  lattice QCD \cite{latticeQCD1,latticeQCD2} and QCD sum rules
  \cite{QCDsumrules}.

\end{enumerate}

We therefore see that the decay $\bsmumu$ is involved in a number of
measurements that are in disagreement with the SM. This raises the
question: assuming that NP is indeed present in $\bsmumu$, what do the
above measurements tell us about it?

Following the announcement of the $R_{K*}$ result, a number of papers
appeared that addressed this question \cite{Capdevila:2017bsm,
  Altmannshofer:2017yso, DAmico:2017mtc, Hiller:2017bzc, Geng:2017svp,
  Ciuchini:2017mik, Celis:2017doq, DiChiara:2017cjq, Sala:2017ihs,
  Ghosh:2017ber}.  The general consensus is that there is a
significant disagreement with the SM, possibly as large as $\sim
6\sigma$, even taking into account the theoretical hadronic
uncertainties \cite{BK*mumuhadunc1,BK*mumuhadunc2,BK*mumuhadunc3}.
These papers generally use a model-independent analysis: $\bsmumu$
transitions are defined via the effective Hamiltonian\footnote{In
  Refs.~\cite{bsmumuNPCPC,bsmumuNPCPV}, it was shown that, when all
  constraints are taken into account, $S$, $P$ and $T$ operators do
  not significantly affect $\BKstarmumu$ (and, by extension, $\bs \to
  \phi \mu^+ \mu^-$) decays.  For this reason only $V$ and $A$
  operators are included in Eq.~(\ref{Heff}). In
  Ref.~\cite{Bardhan:2017xcc}, $T$ operators for both $\bsmumu$ and
  $\bsee$ are considered as a possible explanation of the $R_{K^*}$
  anomaly at low $q^2$.}
\bea
H_{\rm eff} &=& - \frac{\alpha G_F}{\s \pi} V_{tb} V_{ts}^*
      \sum_{a = 9,10} ( C_a O_a + C'_a O'_a ) ~, \nn\\
O_{9(10)} &=& [ {\bar s} \gamma_\mu P_L b ] [ {\bar\mu} \gamma^\mu (\gamma_5) \mu ] ~,
\label{Heff}
\eea
where the $V_{ij}$ are elements of the Cabibbo-Kobayashi-Maskawa (CKM)
matrix. The primed operators are obtained by replacing $L$ with
$R$. If present in $\bsmumu$, NP will contribute to one or more of
these operators. The Wilson coefficients (WCs) $C^{(\prime)}_a$
therefore include both SM and NP contributions. The explanation of
Ref.~\cite{Capdevila:2017bsm} for this discrepancy is that the NP in
$b \to s \mu^+ \mu^-$ satisfies one of three scenarios:
\bea
&{\rm (I)}& ~~~C_9^{\mu\mu}({\rm NP}) < 0 ~, \nn\\
&{\rm (II)}& ~~~C_9^{\mu\mu}({\rm NP}) = - C_{10}^{\mu\mu}({\rm NP}) < 0 ~, \nn\\
&{\rm (III)}& ~~~C_9^{\mu\mu}({\rm NP}) = - C_{9}^{\prime \mu\mu}({\rm NP}) < 0 ~.
\label{bsmumuWCs}
\eea

In the past, numerous models have been proposed that generate the
correct NP contribution to $\bsmumu$ at tree level. A few of them use
scenario (I) above, though most use scenario (II). These models can be
separated into two categories\footnote{New physics from four-quark
  operators can also generate corrections to $C_9$
  \cite{Datta:2013kja}, but they do not lead to lepton universality
  violation and so we not consider them here.}: those containing
leptoquarks (LQs) \cite{CCO,AGC,HS1,GNR,VH,SM,FK,BFK,BKSZ}, and those
with a $Z'$ boson
\cite{CCO,Crivellin:2015lwa,Isidori,dark,Chiang,Virto,GGH,BG,BFG,Perimeter,CDH,SSV,CHMNPR,CMJS,BDW,FNZ,Carmona:2015ena,AQSS,CFL,Hou,CHV,CFV,CFGI,IGG,BdecaysDM,ZMeV,Megias:2017ove,Ahmed:2017vsr}.

We therefore see that there is a wide range of information regarding
the NP in $\bsmumu$, and it is not clear how it is all related. In
Ref.~\cite{RKRDmodels}, it was argued that one has to use
model-independent results carefully, because they may not apply to all
models. To be specific, a particular model may have additional
theoretical or experimental constraints. When these are taken into
account, the results of the model-independent and model-dependent fits
may be significantly different. With this in mind, the purpose of this
paper is to combine the model-independent and model-dependent
analyses, including all the latest measurements, to arrive at a simple
and coherent description of the NP that can explain the data through
its contributions to $\bsmumu$.

We will show the following:
\begin{itemize}

\item Model independent: the NP in $\bsmumu$ follows scenario (I) or
  (II) of Eq.~(\ref{bsmumuWCs}).

\item Model dependent: the simplest NP models are those that involve
  the tree-level exchange of a LQ or a $Z'$. Scenario (II) can arise
  in LQ or $Z'$ models, but scenario (I) is only possible with a $Z'$.

\item Scenario (III) of Eq.~(\ref{bsmumuWCs}) can explain the
  $\bsmumu$ data, but it predicts $R_K = 1$, in disagreement with
  measurement. Furthermore, since it requires an axial-vector coupling
  of the $Z'$, it can only arise in contrived $Z'$ models. For these
  reasons, we exclude it as a possible explanation.

\item In $Z'$ models (i.e., in scenario (I)), there are additional
  constraints from $\bs$-$\bsbar$ mixing and neutrino trident
  production \cite{trident}. A good fit is found only when the
  ${\bar\mu}\mu Z'$ coupling is reasonably (but not too) large. It may
  have an observable effect in a future experiment on neutrino trident
  production.

\item The LQ must be heavy [O(TeV)], but the $Z'$ can be heavy or
  light. For example, we find that the $B$-decay anomalies can be
  explained in $Z'$ models with $M_{Z'} = 10$ GeV or 200 MeV.

\end{itemize}

We begin in Sec.~2 with a description of our method for fitting the
data, including all the latest measurements. The $\bsmumu$ data used
in the fits are given in the Appendix. In Sec.~3 we perform our
model-independent analysis. We turn to the model-dependent analysis in
Sec.~4, separately examining the LQ and $Z'$ models, and making the
connection with the model-independent results. We conclude in Sec.~5.

\section{Fit}

In the following sections, we perform model-independent and
model-dependent analyses of the data. In both cases, we assume that
the NP affects the WCs $C_i$ according to one of three scenarios,
given in Eq.~(\ref{bsmumuWCs}). For each scenario, all observables
are written as functions of the WCs, which contain both SM and
NP contributions and are taken to be real\footnote{The case of
  complex WCs, which can lead to CP-violating effects, is considered
  in Ref.~\cite{bsmumuCPV}.}. Given values of the WCs, we use {\tt
  flavio} \cite{flavio} to calculate the observables
$\mathcal{O}_{th}(C_i)$. Using these, we can compute the $\chi^2$:
\beq
\chi^2(C_i) = (\mathcal{O}_{th}(C_i) -\mathcal{O}_{exp})^T \, \mathcal{C}^{-1} \,
(\mathcal{O}_{th}(C_i) -\mathcal{O}_{exp}) ~,
\eeq
where $\mathcal{O}_{exp}$ are the experimental measurements of the
observables. All available theoretical and experimental correlations
are included in our fit. The total covariance matrix $\mathcal{C}$ is
the sum of the individual theoretical and experimental covariance
matrices, respectively $\mathcal{C}_{th}$ and $\mathcal{C}_{exp}$.  To
obtain $\mathcal{C}_{th}$, we randomly generate all input parameters
and then calculate the observables for these sets of inputs
\cite{flavio}. The uncertainty is then defined by the standard
deviation of the resulting spread in the observable values. In this
way the correlations are generated among the various observables that
share some common parameters \cite{flavio}.  Experimental correlations
are are only available (bin by bin) among the angular observables in
$B \to K^{(*)} \mu^+ \mu^-$ \cite{BK*mumuLHCb2}, and among the angular
observables in $\bs \to \phi \mu^+ \mu^-$ \cite{BsphimumuLHCb2}.

The program {\tt MINUIT}
\cite{James:1975dr,James:2004xla,James:1994vla} is then used to find
the values of the WCs that minimize the $\chi^2$. In this way one can
determine the pull of each scenario, which shows to what extent that
scenario provides a better fit to the data than the SM alone.

There are a number of observables that depend only on $\bsmumu$
transitions. These can clearly be used to constrain NP in $\bsmumu$.
On the other hand, $R_{K^*}$ and $R_K$ also involve $\bsee$
transitions. These can be used to constrain NP in $\bsmumu$ only if
one makes the additional assumption that there is no NP in $\bsee$. We
therefore perform two types of fit. In fit (A), we include only
CP-conserving $\bsmumu$ observables, while in fit (B) we add $R_K$ and
$R_{K^*}$.

The CP-conserving $\bsmumu$ observables are
\begin{enumerate}

\item $B^0 \to K^{*0} \mu^+ \mu^-$: The differential branching ratio
  and the angular observables (see Ref.~\cite{bsmumuCPV} for
  definitions) are measured in various $q^2$ bins. The experimental
  measurements are given in Tables \ref{B0K*mumuBRmeas} and
  \ref{tab:BtoKstar} in the Appendix.

\item $B^+ \to K^{*+} \mu^+ \mu^-$, $B^+ \to K^+ \mu^+ \mu^-$, $B^0
  \to K^0 \mu^+ \mu^-$: The experimental measurements of the
  differential branching ratios of these three decays are given
  respectively in Tables \ref{B+K*mumuBRmeas}, \ref{B+KmumuBRmeas} and
  \ref{B0KmumuBRmeas} in the Appendix.

\item $\bs \to \phi \mu^+ \mu^-$: The differential branching ratio and
  the angular observables are measured in various $q^2$ bins. The
  experimental measurements are given in Tables \ref{BsphimumuBRmeas}
  and \ref{Bsphimumuangmeas} in the Appendix.

\item $B \to X_s \mu^+ \mu^-$: The experimental measurements of the
  differential branching ratio of this decay are given in Table
  \ref{BXsmumuBRmeas} in the Appendix.

\item ${\rm BR}(\bs \to \mu^+ \mu^-) = (2.9 \pm 0.7) \times 10^{-9}$
  \cite{Aaij:2013aka,CMS:2014xfa}.

\end{enumerate}

A comment about the angular observables in $B^0 \to K^{*0} \mu^+
\mu^-$ is in order.  Both LHCb and ATLAS provide measurements of the
$CP$-averaged angular observables $S_i$ as well as the ``optimized''
observables $P_i$, whereas CMS has performed measurements only of the
$P_i$ observables.  In our fits, we have used the measurements of the
$P_i$.  Note that, in Ref.~\cite{Altmannshofer:2017fio}, it was shown
that the best-fit regions and pulls do not change significantly if one
uses the $S_i$ instead of $P_i$ as constraints. Also, we discard the
measurements in $q^2$ bins above 6 $\rm GeV^2$ and below the $J/\psi$
resonance, as the theoretical calculations based on QCD factorization
are not reliable in this region \cite{Beneke:2001at}. In addition, we
discard measurements in bins above the $\psi(2S)$ resonance that are
less than 4 $\rm GeV^2$ wide, as in this region the theoretical
predictions are valid only for $q^2$-integrated observables
\cite{Beylich:2011aq}. LHCb and and ATLAS provide measurements in
different choices of $q^2$ bins. Here we have made sure to use the
data without over-counting.

As noted above, fit (A) includes only the above CP-conserving
$\bsmumu$ observables. However, fit (B) includes $R_{K^*}$ and $R_K$.
To perform fit (B), we followed the same strategy as in the recent
global analysis of Ref.~\cite{Capdevila:2017bsm}, namely we
simultaneously included both ${\cal B}(B^0 \to K^{(*)0} \mu^+\mu^-)$
and $R_K^{(*)}$ in the fit. Since these observables are expected to be
correlated, one might worry about overcounting. However, we found very
similar results when ${\cal B}(B^0 \to K^{(*)0} \mu^+\mu^-)$ for the
low-$q^2$ bins were removed from the fit.

Fits (A) and (B) are used in both the model-independent and
model-dependent analyses. However, a particular model may receive
further constraints from its contributions to other observables, such
as $\bsnunubar$, $\bs$-$\bsbar$ mixing and neutrino trident
production. These additional constraints will be taken into account in
the model-dependent fits.

\section{Model-independent analysis}

\subsection{\bf Fit (A)}

We begin by applying fit (A), which involves only the CP-conserving
$\bsmumu$ observables, to the three scenarios. The results are shown
in Table \ref{micouplings1}. All scenarios can explain the data, with
  pulls of roughly 5.

\begin{table}[htb]
\begin{center}
\begin{tabular}{|c|c|c|} \hline
Scenario & WC & pull  \\
\hline
(I) $C_9^{\mu\mu}({\rm NP}) $
         & $ -1.20 \pm 0.20 $  & 5.0  \\
\hline
(II) $C_9^{\mu\mu}({\rm NP})  = -C_{10}^{\mu\mu}({\rm NP}) $
         & $ -0.62 \pm 0.14 $  & 4.6  \\
\hline
(III) $C_9^{\mu\mu}({\rm NP})  = -C_{9}^{'{\mu\mu}}({\rm NP}) $
         & $ -1.10 \pm 0.18 $  & 5.2  \\
\hline
\end{tabular}
\end{center}
\caption{Model-independent scenarios: best-fit values of the WCs
  (taken to be real), as well as the pull = $\sqrt{\chi^2_{SM} -
    \chi^2_{SM + NP}}$ for fit (A) (only CP-conserving $\bsmumu$
  observables). For each case there are 112 degrees of freedom.
\label{micouplings1}}
\end{table}

\subsection{\bf Fit (B)}
\label{FitB}

We now examine how the three scenarios fare when confronted with the
$R_{K^*}$ and $R_K$ data. One way to take into account the constraints
from $R_{K^*}$ and $R_K$ is to incorporate them into the fit [fit
  (B)]. The results for the three scenarios are shown in Table
\ref{micouplings2}. In comparing fits (A) and (B), we note the
following:
\begin{itemize}

\item The addition of $R_{K^*}$ and $R_K$ to the fit has led to a
  substantial quantitative increase in the disagreement with the SM.
  In fit (A) the average pull is 4.9, while in (B) it is 5.8.

\item The increase in the pull is 0.9, 1.3 and 0.4 for scenarios (I),
  (II) and (III), respectively. In fit (A), scenario (III) has the
  largest pull, while in (B) it is the smallest. Still, with a pull of
  5.6, scenario (III) appears to be a viable candidate for explaining
  the $\bsmumu$ anomalies.

\end{itemize}

\begin{table}[htb]
\begin{center}
\begin{tabular}{|c|c|c|} \hline
Scenario & WC & pull  \\
\hline
(I) $C_9^{\mu\mu}({\rm NP}) $
         & $ -1.25 \pm 0.19 $  & 5.9  \\
\hline
(II) $C_9^{\mu\mu}({\rm NP})  = -C_{10}^{\mu\mu}({\rm NP}) $
         & $ -0.68 \pm 0.12 $  & 5.9  \\
\hline
(III) $C_9^{\mu\mu}({\rm NP})  = -C_{9}^{'{\mu\mu}}({\rm NP}) $
         & $ -1.11 \pm 0.17 $  & 5.6  \\
\hline
\end{tabular}
\end{center}
\caption{Model-independent scenarios: best-fit values of the WCs
  (taken to be real), as well as the pull = $\sqrt{\chi^2_{SM} -
    \chi^2_{SM + NP}}$ for fit (B) (CP-conserving $\bsmumu$ observables
  $+$ $R_{K^*}$ and $R_K$). For each case there are 115 degrees of
  freedom.
\label{micouplings2}}
\end{table}

\subsection{\bf \boldmath Predictions of $R_{K^*}$ and $R_K$}
\label{predictRK(*)}

\begin{figure*}[htb]
\centering
\includegraphics[height=6.0cm,width=7.5cm]{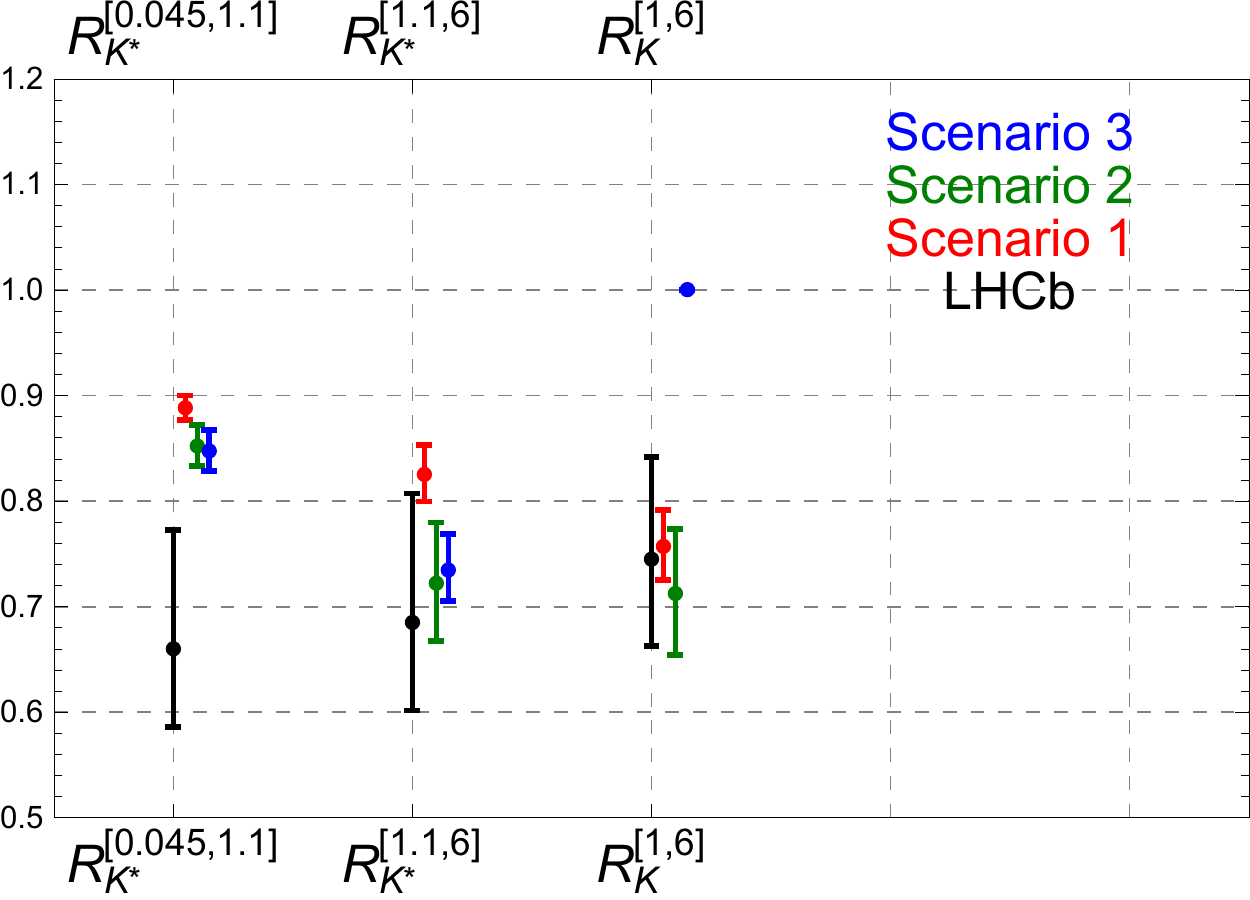}
\caption{Comparison of the experimental measurements of $R_K$ and
  $R_{K^*}$ with the predictions of the three scenarios.}
\label{fig:rks-rk-I-III}
\end{figure*}

Another way to include considerations of $R_{K^*}$ and $R_K$ is simply
to take the preferred WCs from Table \ref{micouplings1} and predict
the allowed values of $R_{K^*}$ and $R_K$ in the three scenarios. The
results are shown in Fig.~\ref{fig:rks-rk-I-III}.

The first thing one sees is that none of the three scenarios predict a
value for $R_{K^*}$ in the low-$q^2$ bin that is in agreement (within
$1\sigma$) with the experimental measurement [Eq.~(\ref{RK*meas})]. In
the SM, in this $q^2$ region, the decay $\bsll$ is dominated by the
photon contribution, parametrized by the WC $C_7$
\cite{RK*theory}. Since the photon coupling is lepton flavor
universal, it is only threshold effects, with $m_\mu \ne m_e$, that
lead to $R_{K^*}^\SM \simeq 0.93$ \cite{flavio}. It is difficult to
find NP that can compete with the photon contribution and
significantly change $R_{K^*}$ from its SM prediction. On the other
hand, the discrepancy between the measurement and the predictions is
only at the level of approximately $1.5\sigma$, which is not
worrisome.

The predictions for the remaining measurements agree with the
experimental values, with one glaring exception. Scenario (III)
predicts $R_K = 1$, as in the SM. This is in disagreement with the
measurement [Eq.~(\ref{RKexpt})].

As was shown in Sec.~\ref{FitB}, when $R_{K^*}$ and $R_K$ are included
in the fit [fit (B)], the overall result with scenario (III) is good
(a pull of 5.6). This scenario can therefore be considered a possible
explanation for the $B$-decay anomalies. (Indeed, this is the
conclusion of Ref.~\cite{Capdevila:2017bsm}.) However, in our opinion,
this is not sufficient. As we saw above, scenario (III) predicts a
value for $R_K$ that is in striking disagreement with the measurement.
Furthermore, $R_K$ is a clean observable, i.e., it has very little
theoretical uncertainty, so theoretical error cannot be a reason for
the disagreement. The only reason fit (B) gives a good fit is that the
$R_K$ measurement is only one of many, so its effect is
diminished. However, we feel that this is misleading: given its clear
failure to explain the measured value of $R_K$, scenario (III) should
be considered as strongly disfavored, compared to scenarios (I) and
(II).

$R_{K^*}$ and $R_K$ have been measured in the region of $q^2 \le 6$
$\rm GeV^2$. It is likely that these observables will also be measured
in the region $15 \le q^2 \le 22$ $\rm GeV^2$.  Below we present the
predictions of the three scenarios for $R_{K^*}$ and $R_K$ in this
high-$q^2$ bin:
\bea
R_{K^*} &~=~& 0.76 \pm 0.03~({\rm I}) ~~,~~~~ 0.71 \pm 0.06~({\rm II}) ~~,~~~~ 0.68 \pm 0.04~({\rm III}) ~, \nn\\
R_K &~=~& 0.76 \pm 0.03~({\rm I}) ~~,~~~~ 0.72 \pm 0.05~({\rm II}) ~`,~~~~ 1.0~({\rm III}) ~.
\eea

\section{Model-dependent analysis}

The simplest NP models one can construct that explain the $B$
anomalies involve the tree-level exchange of a new particle. This
particle can be either a leptoquark or a $Z'$ boson. Below we examine
the properties of such NP models required for them to account for the
$\bsmumu$ decays.

\subsection{\bf Leptoquarks}

LQ models were studied in detail in Ref.~\cite{bsmumuCPV}. It was
found that, of the ten LQ models that couple to SM particles through
dimension $\le 4$ operators, only three can explain the $\bsmumu$
data. They are: a scalar isotriplet with $Y = 1/3$, a vector
isosinglet with $Y = -2/3$, and a vector isotriplet with $Y = -2/3$.
These are denoted $S_3$, $U_1$ and $U_3$, respectively
\cite{Sakakietal}. As far as the $\bsmumu$ processes are concerned,
the models all have $C_9^{\mu\mu}({\rm NP}) = - C_{10}^{\mu\mu}({\rm
  NP})$, and so are equivalent. That is, all LQ models fall within
scenario (II) of Eq.~(\ref{bsmumuWCs}).

The $S_3$, $U_1$ and $U_3$ LQ models all contribute differently to $b
\to s \nu_\mu {\bar\nu}_\mu$ decays, so that, in principle, they can
be distinguished.  However, it was shown in Ref.~\cite{bsmumuCPV} that
the present constraints from $B \to K^{(*)} \nu {\bar\nu}$ are far
weaker than those from $\bsmumu$ processes, so that the current
$\bsnunubar$ data cannot be used to distinguish the three LQ models.
(This said, this conclusion can be evaded if the LQs couple to other
leptons, see Ref.~\cite{RKRDmodels} for an example.)

The bottom line is that there is effectively only a single LQ model
that can explain the $B$-decay anomalies, and it is of type scenario
(II). In order to determine the value of the WC required to reproduce
the $\bsmumu$ data, a fit to this data is required, including all
other processes to which this type of NP contributes. In this case,
the only additional process is $b \to s \nu_\mu {\bar\nu}_\mu$, which
does not furnish any additional constraints. The allowed value of the
WC is therefore the same as that found in the model-independent fit,
in Table \ref{micouplings1} or \ref{micouplings2}.

This $\bsmumu$ WC is generated by the tree-level exchange of a LQ.
Thus,
\beq
C_9^{\mu\mu}({\rm NP}) \propto \frac{g_L^{b\mu} g_L^{b\mu}}{M_{\rm LQ}^2} ~,
\eeq
where $g_L^{b\mu}$ and $g_L^{s\mu}$ are the couplings of the LQ (taken
to be real), and $M_{\rm LQ}$ its mass. Direct searches constrain
$M_{\rm LQ} > 640$ GeV \cite{LQmasslimits}.

\subsection{\bf \boldmath $Z'$ bosons}
\label{Z'bosons}

In the previous subsection, we saw that LQ models are all of type
scenario (II).  This implies that scenarios (I) and (III) can
only occur within $Z'$ models. Is this possible? The four-fermion
$\bsmumu$ operators required within the four scenarios are as follows:
\bea
&{\rm (I)}& ~~~[ {\bar s} \gamma_\mu P_L b ] [ {\bar\mu} \gamma^\mu \mu ] ~, \nn\\
&{\rm (II)}& ~~~[ {\bar s} \gamma_\mu P_L b ] [ {\bar\mu} \gamma^\mu P_L \mu ] ~, \nn\\
&{\rm (III)}& ~~~[ {\bar s} \gamma_\mu \gamma_5 b ] [ {\bar\mu} \gamma^\mu \mu ] ~.
\eea
Scenarios (I) and (II) are clearly allowed. They require the $Z'$ to
couple vectorially to ${\bar s}_L b_L$ and ${\bar\mu} \mu$ or
${\bar\mu}_L \mu_L$. It is quite natural for gauge bosons to couple
vectorially, so it is easy to construct models which lead to scenario
(I) or (II). On the other hand, scenario (III) requires that the $Z'$
couple axial-vectorially to ${\bar s} b$. This is much less
natural. It is possible to arrange this, but it requires a rather
contrived model (e.g., see Ref.~\cite{Capdevila:2017bsm}).
Furthermore, we have already seen that scenario (III) is strongly
disfavored by the $R_K$ measurement. In light of all this, we
therefore exclude scenario (III) as a realistic explanation of the
$B$-decay anomalies.

The conclusion is that, when model-independent and model-dependent
considerations are combined, only scenarios (I) and (II) are possible
as explanations of the $B$-decay anomalies. Furthermore, while
scenario (II) can be realized with a LQ or $Z'$ model, scenario (I)
can only be due to $Z'$ exchange.

Since the $Z'$ couples to two left-handed quarks, it must transform as
a singlet or triplet of $SU(2)_L$. The triplet option has been
considered in
Refs.~\cite{CCO,Crivellin:2015lwa,Isidori,dark,Chiang,Virto}. (In this
case, there is also a $W'$ that can contribute to ${\bar B} \to
D^{(*)+} \tau^- {\bar\nu}_\tau$ \cite{RKRD}, another decay whose
measurement exhibits a discrepancy with the SM
\cite{RD_BaBar,RD_Belle,RD_LHCb}.) Alternatively, if the $Z'$ is a
singlet of $SU(2)_L$, it must be the gauge boson associated with an
extra $U(1)'$. Numerous models of this type have been proposed, see
Refs.~\cite{GGH,BG,BFG,Perimeter,CDH,SSV,CHMNPR,CMJS,BDW,FNZ,Carmona:2015ena,AQSS,CFL,Hou,
  CHV,CFV,CFGI,IGG,BdecaysDM,ZMeV,Megias:2017ove,Ahmed:2017vsr}.

The vast majority of $Z'$ models that have been proposed assume a
heavy $Z'$, $M_{Z'} =$ O(TeV). This option is examined in
Sec.~\ref{heavyzp}.  However, we also note that the $Z'$ can be
light. The cases of $M_{Z'} = 10$ GeV or 200 MeV are considered in
Sec.~\ref{lightzp}.

\subsubsection{\bf \boldmath Heavy $Z'$}
\label{heavyzp}

In order to determine the properties of $Z'$ models that explain the
$\bsmumu$ data, one cannot simply perform fits (A) or (B) -- important
constraints from other observables must be taken into account. Since
the $Z'$ model is of the type scenario (I) or (II), we can write
\bea
\Delta {\cal L}_{Z'} & = & J^\mu Z'_\mu ~, \nn\\
{\rm where} \qquad J^\mu & = &
g_{L}^{\mu \mu} \, \bar L \gamma^{\mu} P_L L + g_{R}^{\mu \mu} \, \bar \mu \gamma^{\mu} P_R \mu
+ g_L^{bs} \, {\bar\psi}_{q2} \gamma^{\mu} P_L \psi_{q3} + h.c.
\label{Z'couplings}
\eea
Here $\psi_{qi}$ is the quark doublet of the $i^{th}$ generation, and
$L=(\nu_\mu , \mu)^T$.  We have
\bea
{\hbox{scenario (I)}} &~~:~~& g_R^{\mu\mu} = g_L^{\mu\mu} ~, \nn\\
{\hbox{scenario (II)}} &~~:~~& g_R^{\mu\mu} = 0 ~.
\label{scenariodefs}
\eea
When the heavy $Z'$ is integrated out, we obtain the following
effective Lagrangian containing 4-fermion operators:
\bea
{\cal L}_{Z'}^{eff} = -\frac{1}{2 M_{Z'}^2} J_\mu J^\mu
& \supset & -\frac{g_L^{bs}}{M_{Z'}^2} (\bar s \gamma^{\mu} P_L b) (\bar \mu \gamma^{\mu}
(g_{L}^{\mu \mu}P_L + g_{R}^{\mu \mu}P_R) \mu)
- \frac{(g_L^{bs})^2}{2 M_{Z'}^2} (\bar s \gamma^{\mu} P_L b) (\bar s \gamma^{\mu}
P_L b) \nn\\
&& \hskip0.5truecm
-~\frac{g_{L}^{\mu \mu}}{M_{Z'}^2} (\bar \mu \gamma^{\mu} (g_{L}^{\mu \mu}P_L +
g_{R}^{\mu \mu}P_R) \mu) ({\bar \nu}_\mu \gamma^{\mu} P_L \nu_\mu) ~.
\label{Z'4fermi}
\eea
The first 4-fermion operator is relevant for $\bsmumu$ transitions,
the second operator contributes to $\bs$-$\bsbar$ mixing, and the
third operator contributes to neutrino trident production.

$\bullet$ {\bf \boldmath $\bs$-$\bsbar$ mixing:}

The formalism leading to the constraint on $g_L^{bs}$ from
$\bs$-$\bsbar$ mixing is given in Ref.~\cite{bsmumuCPV}. We do not
repeat it here. The one thing to keep in mind is that
Ref.~\cite{bsmumuCPV} considered a complex $g_L^{bs}$, while here it
is taken to be real.

$\bullet$ {\bf Neutrino trident production:}

The production of $\mu^+\mu^-$ pairs in neutrino-nucleus scattering,
$\nu_\mu N \to \nu_\mu N \mu^+ \mu^-$ (neutrino trident production),
is a powerful probe of new-physics models \cite{trident}. The heavy
$Z'$ contribution to this process is also given in
Ref.~\cite{bsmumuCPV}. However, there only scenario (II)
($g_R^{\mu\mu} = 0$) is considered. Allowing for a nonzero
$g_R^{\mu\mu} $, one obtains the following: the theoretical prediction
for the cross section is
\beq
 \left. { \sigma_\text{SM+NP} \over \sigma_\text{SM} } \right|_{\nu N \to \nu N \mu^+ \mu^-}
=
\frac{1}{1+(1+4s_W^2)^2} \left [ \left ( 1 + \frac{v^2g_L^{\mu \mu}(g_L^{\mu \mu} - g_R^{\mu \mu})}{M_{{Z}^{'}}^2}  \right )^2
+ \left ( 1 +4 s_W^2 +  \frac{v^2 g_L^{\mu \mu}(g_L^{\mu \mu} + g_R^{\mu \mu})}{M_{{Z}^{'}}^2}  \right )^2 \right ] ~.
\label{tridentHeavy}
\eeq
This is to be compared with the experimental measurement \cite{CCFR}:
\beq
 \left. { \sigma_\text{exp.} \over \sigma_\text{SM} } \right|_{\nu N \to \nu N \mu^+ \mu^-} = 0.82 \pm 0.28 ~.
\eeq
Using Eq.~(\ref{scenariodefs}), this comparison provides an upper
limit on $(g_L^{\mu\mu})^2/M_{Z'}^2$.  For $M_{Z'}=1 $ TeV and $v =
246$ GeV, we obtain the following $1\sigma$ upper bound on the
coupling:
\bea
{\rm (I)} :  ~~~~~ |g_L^{\mu\mu}| \le 0.99 ~, \nn\\
{\rm (II)} :  ~~~~~ |g_L^{\mu\mu}| \le 1.38 ~.
\label{tridentconstraints}
\eea

$\bullet$ {\bf \boldmath $\bsmumu$:}

The couplings $g_L^{bs}$ and $g_{L,R}^{\mu\mu}$ are all involved in $\bsmumu$:
\bea
C_9^{\mu \mu}({\rm NP}) &=& -\left[ \frac{\pi}{\sqrt 2 G_F \alpha V_{tb} V_{ts}^*} \right ] \,
\frac{g_L^{bs} (g_L^{\mu \mu} + g_R^{\mu \mu}) }{M_{Z'}^2} ~, \nn\\
C_{10}^{\mu \mu}({\rm NP}) &=& \left[ \frac{\pi}{\sqrt 2 G_F \alpha V_{tb} V_{ts}^*} \right ] \,
\frac{g_L^{bs} (g_L^{\mu \mu} - g_R^{\mu \mu}) }{M_{Z'}^2} ~.
\label{WCNPmodel}
\eea
We see that any analysis of $Z'$ models must include the constraints
from $\bs$-$\bsbar$ mixing and neutrino trident production. And this
applies to scenario (I), which, though supposedly model-independent,
is related to $Z'$ models.

The results of fits (A) and (B) are given in Tables
\ref{tab:heavyzprime_fitA} and \ref{tab:heavyzprime_fitB},
respectively. These illustrate quite clearly the connection between
the model-independent and model-dependent approaches. From the
model-independent point of view, in order to explain the experimental
data, the NP WC must take a certain value (given in Tables
\ref{micouplings1} and \ref{micouplings2}). However, from the
model-dependent point of view, this WC is proportional to the product
$g_L^{bs} g_L^{\mu \mu}$ [Eq.~(\ref{WCNPmodel}), using
  Eq.~(\ref{scenariodefs})], and these individual couplings have
additional constraints from other processes. $g_L^{\mu \mu}$ is
constrained by neutrino trident production
[Eq.~(\ref{tridentconstraints})]. Now, if $g_L^{\mu \mu}$ is small,
$g_L^{bs}$ must be large in order to reproduce the required
WC. However, a large $g_L^{bs}$ is in conflict with the constraint
from $\bs$-$\bsbar$ mixing, resulting in a poorer fit (i.e., a smaller
pull). On the other hand, if $g_L^{\mu \mu}$ is large (but still
consistent with Eq.~(\ref{tridentconstraints})), $g_L^{bs}$ can be
small, so that the $\bs$-$\bsbar$ mixing constraint is less
important. In this case, a good fit (i.e., a large pull) is
possible. Indeed, for large enough $g_L^{\mu \mu}$, one simply
reproduces the model-independent result. For both fits (A) and (B), we
find that this is the case for $g_L^{\mu \mu} \ge 0.4$. The conclusion
is that, if the NP is a $Z'$, the coupling $g_L^{\mu \mu}$ has to be
reasonably big. Its effect may be observable in a future experiment on
neutrino trident production.

\begin{table}[htb]
\centering
\begin{tabular}{|c|c|c|} \hline
 & $M_{Z'} = 1$ TeV &  \\ \hline
$g_L^{\mu \mu}$ & $Z'$ (I): $g_L^{bs}$ $\times 10^{3}$ & pull  \\
\hline
0.01
         &$-2.6  \pm 1.9 $ & 1.0  \\
         \hline
0.05
         &$-4.2  \pm 1.1 $ & 2.8  \\
         \hline
0.1
         &$-4.6  \pm 0.9 $ &4.0   \\
         \hline
0.2
         &$-3.8  \pm 0.7 $ &4.8  \\
         \hline
0.4
         &$-2.2  \pm 0.4 $ &5.0 \\
         \hline
0.5
         & $-1.8  \pm 0.3 $ & 5.0 \\
\hline
\hline
\end{tabular}
~~~~~~~~~
\begin{tabular}{|c|c|c|} \hline
 & $M_{Z'} = 1$ TeV &  \\ \hline
$g_L^{\mu \mu}$ & $Z'$ (II): $g_L^{bs}$ $\times 10^{3}$ & pull  \\
\hline
0.01
         &$-2.4 \pm 1.9 $& 1.0   \\
         \hline
0.05
         &$-4.0 \pm 1.1 $ & 2.8 \\
         \hline
0.1
         & $-3.6 \pm 0.8 $ &3.6   \\
         \hline
0.2
         & $-3.8 \pm 0.8 $ &4.3   \\
         \hline
0.4
         & $-2.3 \pm 0.5 $ &4.6  \\
         \hline
0.5
         &  $-1.9 \pm 0.4 $ & 4.6   \\
\hline
\hline
\end{tabular}
\caption{$Z'$ model (scenario (I) : left, scenario (II) : right):
  best-fit value of $g_L^{bs}$, and the pull=$\sqrt{\chi^2_{SM}
    -\chi^2_{SM + NP} }$ for fit (A) (only CP-conserving $\bsmumu$
  observables), for various values of $g_L^{\mu \mu}$.
\label{tab:heavyzprime_fitA}}
\end{table}

\begin{table}[htb]
\centering
\begin{tabular}{|c|c|c|} \hline
 & $M_{Z'} = 1$ TeV &  \\ \hline
$g_L^{\mu \mu}$ & $Z'$ (I): $g_L^{bs}$ $\times 10^{3}$ & pull  \\
\hline
0.01
         &$-3.0  \pm 1.6 $ & 1.4  \\
         \hline
0.05
         &$-4.8  \pm 1.0 $ & 2.8  \\
         \hline
0.1
         &$-5.2  \pm 0.8 $ & 4.5  \\
         \hline
0.2
         &$-4.2  \pm 0.6 $ & 5.7  \\
         \hline
0.4
         &$-2.4  \pm 0.4 $ & 5.9\\
         \hline
0.5
         & $-1.9  \pm 0.3 $ & 5.9 \\
\hline
\hline
\end{tabular}
~~~~~~~~~
\begin{tabular}{|c|c|c|} \hline
 & $M_{Z'} = 1$ TeV &  \\ \hline
$g_L^{\mu \mu}$ & $Z'$ (II): $g_L^{bs}$ $\times 10^{3}$ & pull  \\
\hline
0.01
         &$-3.0 \pm 1.6 $& 1.4   \\
         \hline
0.05
         &$-4.8 \pm 1.0 $ & 2.8 \\
         \hline
0.1
         & $-5.2 \pm 0.8 $ &4.5   \\
         \hline
0.2
         & $-4.4 \pm 0.7 $ & 5.6   \\
         \hline
0.4
         & $-2.5 \pm 0.4 $ &5.9  \\
         \hline
0.5
         &  $-2.1 \pm 0.4 $ & 5.9   \\
\hline
\hline
\end{tabular}
\caption{$Z'$ model (scenario (I) : left, scenario (II) : right):
  best-fit value of $g_L^{bs}$, and the pull=$\sqrt{\chi^2_{SM}
    -\chi^2_{SM + NP} }$ for fit (B) (CP-conserving $\bsmumu$ observables
  $+$ $R_{K^*}$ and $R_K$), for various values of $g_L^{\mu \mu}$.
\label{tab:heavyzprime_fitB}}
\end{table}

\subsubsection{\bf \boldmath Light $Z'$}
\label{lightzp}

An interesting possibility to consider is a light $\Z$. If the $\Z$
mass is between $m_B$ and $2 m_\mu$, then, if it is narrow, one can
observe this state as a resonance in the dimuon invariant mass. Since
no such state has been observed, we consider the mass ranges $m_{\Z} >
m_B$ and $m_{\Z} < 2 m_\mu$. A $ \Z$ in the first mass range may have
implications for dark matter phenomenology \cite{BdecaysDM}, while a
$\Z$ in the second mass range could explain the muon $g-2$ measurement
and have implications for nonstandard neutrino interactions
\cite{ZMeV}. For the first mass range we consider $M_{Z'} = 10 $ GeV
and refer to this as the GeV $\Z$ model, while in the second range we
consider $M_{Z'} = 200 $ MeV and call it the MeV $\Z$
model\footnote{After the $R_K^*$ measurement was announced, a GeV $\Z$
  model was considered in Ref.~\cite{Sala:2017ihs} and an MeV $\Z$
  model in Ref.~\cite{Ghosh:2017ber}.}

For the MeV $\Z$ model, we assume there is a flavor-changing ${\bar
  s}b\Z$ vertex whose form is taken to be
\beq
F(q^2) \, \bar{s} \gamma^{\mu}P_L  b \, Z^\prime_{\mu} ~.
\eeq
The form factor $F(q^2)$ is expanded for the momentum transfer $q^2
\ll m_B^2$ as
\bea
F(q^2) & = & a^{bs}_L + g^{bs}_L \frac{q^2}{m_B^2} + \ldots\,,\
\eea
where $m_B$ is the $B$-meson mass. For the GeV $\Z$ model there is no
form factor, and the ${\bar s}b\Z$ vertex is taken to be fixed at
$a^{bs}_L$ for all $q^2$.

In the MeV $\Z$ model, assuming the $\Z$ couples to neutrinos, the
leading-order term $a^{bs}_L$ is constrained by $B \to K \nu \bar\nu$
to be smaller than $10^{-9}$. To explain the $\bsmumu$ anomalies, we
then require the $\Z$ to have a large coupling to muons, which is
inconsistent with data \cite{ZMeV}. We therefore neglect $a^{bs}_L$
and keep only $g^{bs}_L$. (If the $\Z$ does not couple to neutrinos
then this constraint does not apply.) In the GeV $\Z$ model $a^{bs}_L$
is present, so here we neglect $g^{bs}_L$.

The matrix  elements for the various processes are then
\bea
M_{\bsmumu}
& = &
-\frac{ F(q^2)}{q^2- M_{Z'}^2} (\bar s \gamma^{\mu} P_L b)
(\bar \mu \gamma^{\mu} (g_{L}^{\mu \mu}P_L + g_{R}^{\mu \mu}P_R) \mu) ~, \nn\\
M_{{B_s}mix} & = &
- \frac{F(q^2)^2}
  {2 q^2- 2 M_{Z'}^2} (\bar s \gamma^{\mu} P_L b) (\bar s
\gamma^{\mu} P_L b)  \left[ 1- \frac{5}{8} \frac{m_b^2}{m_{\Z}^2} \right] ~,    \nn\\
M_{trident} & = &
-~\frac{g_L^{\mu \mu}}{q^2 -M_{Z'}^2} (\bar \mu \gamma^{\mu} (g_{L}^{\mu \mu}P_L +
g_{R}^{\mu \mu}P_R) \mu) ({\bar \nu}_\mu \gamma^{\mu} P_L \nu_\mu) ~,
\eea
where we have used Ref.~\cite{tandean} for $\bs$-$\bsbar$ mixing.  In
$M_{\bsmumu}$ there is an additional contribution from the
longitudinal $\Z$ for the axial leptonic current that is $\sim m_\mu
m_b / m_{\Z}^2$. For the GeV $\Z$ model this term can be neglected.
However, for the MeV $\Z$ model this term is sizeable, and so for this
case we only consider scenario I with a vectorial leptonic current.
As usual, we assume the $\Z$ does not couple to electrons, so that
${\cal B}(B^+ \to K^+ e^+ e^-)$ is described by the SM, while ${\cal
  B}(B^+ \to K^+ \mu^+ \mu^-)$ is modified by NP.

$\bullet$ {\bf \boldmath $\bs$-$\bsbar$ mixing:}

The measurement of $\bs$-$\bsbar$ mixing gives a constraint on the
product of couplings and the form factor. For the MeV $\Z$ model, as
the form factor at $q^2 \sim m_B^2$ is not known, we fit $g_L^{bs}$
only from the $\bsmumu$ data, while for the GeV $\Z$ model, where the
form factor is unity, the mixing is used to obtain a constraint on
$a_L^{bs}$.

$\bullet$ {\bf Neutrino trident production:}

The coupling $g^{\mu\mu}$ is constrained by neutrino trident
production. For the MeV $\Z$ model, Eq.~\ref{tridentHeavy} is no
longer valid -- instead we use the constraints from
Ref.~\cite{trident}.  In this reference only scenario (I)
($g_R^{\mu\mu} = g_L^{\mu \mu}$) is considered.  There are other
constraints that the MeV $\Z$ model must satisfy; these are discussed
in Ref.~\cite{Farzan}. All these constraints are consistent with the
constraint obtained from neutrino trident production.

$\bullet$ {\bf \boldmath $\bsmumu$:}

For $\bsmumu$ we have
\bea
C_9^{\mu \mu}({\rm NP}) &=& \left[ \frac{\pi}{\sqrt 2 G_F \alpha V_{tb} V_{ts}^*} \right ] \,
\frac{\left ( a_L^{bs} + g_L^{bs} (q^2 / m_{B_s}^2) \right ) (g_L^{\mu \mu} + g_R^{\mu \mu})} {q^2 - M_{Z'}^2} ~, \nn\\
C_{10}^{\mu \mu}({\rm NP}) &=& -\left[ \frac{\pi}{\sqrt 2 G_F \alpha V_{tb} V_{ts}^*} \right ] \,
\frac{\left ( a_L^{bs} + g_L^{bs} (q^2 / m_{B_s}^2) \right ) (g_L^{\mu \mu} - g_R^{\mu \mu})} {q^2 - M_{Z'}^2} ~.
\eea
Interestingly, here the WCs are $q^2$-dependent.

Using these WCs, we perform a fit to the data.  We scan the parameter
space of $g_{bs}$ and $g_{\mu\mu}$ for values that are consistent with
all experimental measurements.  For the MeV $\Z$ model, the form
factor is not known in the high-$q^2$ region, and so one can fit only
to the low-$q^2$ bins. However, we have checked that the fit does not
change much if we use the above form factor for all $q^2$ bins. For
both the MeV and GeV $Z'$ we find that, in fact, it is possible to
explain the $B$-decay anomalies with pulls that are almost as good as
in the case of a heavy $Z'$.

For the MeV $\Z$ model, the best fit has a pull of 4.4, and is found
for the product of couplings $g_L^{bs} g_L^{\mu \mu} \sim 21 \times
10^{-9}$. Taking $g_L^{\mu \mu} \sim 10^{-3}$ from the neutrino
trident constraint, one obtains $ g_L^{bs} \sim 2.1 \times 10^{-5}$,
which is consistent with constraints from $B \to K \nu \bar{\nu}$
\cite{ZMeV}.  The results for the GeV $\Z$ model are shown in Table
\ref{tab:GeVzprime} for fit (A). The best fit has a pull of 4.2
(scenario (I)) or 4.5 (scenario (II)).

\begin{table}[htb]
\centering
\begin{tabular}{|c|c|c|} \hline
 & $M_{Z'} = 10$ GeV &  \\ \hline
$g_L^{\mu \mu}$ $\times 10^{2}$ & $Z'$ (I): $g_L^{bs}$ $\times 10^{6}$ & pull  \\
\hline
0.05
         &$-36.3  \pm 10.2 $ & 2.6  \\
         \hline

0.1
         &$-37.6  \pm 8.5 $ & 3.6  \\
         \hline
0.3
         &$-20.2  \pm 4.6 $ &4.1   \\
         \hline
0.6
         &$-10.3  \pm 2.3 $ &4.2  \\
         \hline
0.9
         &$-6.9  \pm 1.6 $ &4.2 \\
         \hline
1.2
         & $-5.2  \pm 1.2 $ & 4.2 \\
\hline
\hline
\end{tabular}
~~~~~~~~~
\begin{tabular}{|c|c|c|} \hline
 & $M_{Z'} = 10$ GeV &  \\ \hline
$g_L^{\mu \mu}$ $\times 10^2$ & $Z'$ (II): $g_L^{bs}$ $\times 10^{6}$ & pull  \\
\hline
0.05
         &$-35.4 \pm 11.0 $& 2.8   \\
         \hline
0.1
         &$-38.7 \pm 9.0 $ & 3.4 \\
         \hline
0.3
         & $-27.0 \pm 6.2 $ &4.3   \\
         \hline
0.6
         & $-14.4 \pm 3.6 $ &4.5   \\
         \hline
0.9
         & $-9.6 \pm 2.3 $ &4.5  \\
         \hline
1.2
         &  $-7.2 \pm 1.8 $ & 4.5   \\
\hline
\hline
\end{tabular}
\caption{GeV $Z'$ model (scenario (I) : left, scenario (II) : right):
  best-fit value of $g_L^{bs}$, and the pull=$\sqrt{\chi^2_{SM}
    -\chi^2_{SM + NP} }$ in fit (A), for various values of $g_L^{\mu
    \mu}$.
\label{tab:GeVzprime}}
\end{table}

As noted in the discussion about Fig.~\ref{fig:rks-rk-I-III}, the
value of $R_{K^*}$ in the low-$q^2$ bin ($0.045 \le q^2 \le 1.1 ~{\rm
  GeV}^2$) is dominated by the SM photon contribution. Heavy NP cannot
significantly affect this, and so cannot much improve the discrepancy
between the measurement and the SM prediction of $R_{K^*}$ in this
bin. On the other hand, since the WCs are $q^2$-dependent in
light-$Z'$ models, in principal they could have a large effect on this
value of $R_{K^*}$. Unfortunately, for $M_{Z'} = 10$ GeV and 200 MeV,
we find that the prediction for $R_{K^*}$ in the low-$q^2$ bin is
little changed from that of the SM. However, this might not hold in a
different version of a light $Z'$ model (for example, see
Ref.~\cite{Ghosh:2017ber}).

\section{Conclusions}

Following the announcement of the measurement of $R_{K^*}$
\cite{RK*expt}, a flurry of papers appeared
\cite{Capdevila:2017bsm,Altmannshofer:2017yso,DAmico:2017mtc,Hiller:2017bzc,Geng:2017svp,Ciuchini:2017mik,Celis:2017doq,DiChiara:2017cjq,
  Sala:2017ihs,Ghosh:2017ber} discussing how to explain the result and
what it implies for new physics. Most papers adopted a
model-independent approach, while a few focused on particular
models. The main purpose of the present paper is to show that
additional information about the NP is available if one combines the
model-independent and model-dependent analyses.

To be specific, the general preference was for NP in $\bsmumu$
transitions (although some papers considered the possibility of NP in
both $\bsmumu$ and $\bsee$). Several model-independent studies pointed
out that the $\bsmumu$ anomalies can be explained if (I)
$C_9^{\mu\mu}({\rm NP}) < 0$ or (II) $C_9^{\mu\mu}({\rm NP}) = -
C_{10}^{\mu\mu}({\rm NP}) < 0$. We agree with this observation. Now,
the simplest NP models involve the tree-level exchange of a leptoquark
(LQ) or a $Z'$ boson. A number of different LQ models have previously
been proposed, but we point out that, as far as the $\bsmumu$
processes are concerned, all viable models have $C_9^{\mu\mu}({\rm
  NP}) = - C_{10}^{\mu\mu}({\rm NP})$, and so are equivalent. That is,
there is effectively a single LQ model, and it falls within scenario
(II).

The key point is that, although scenario (II) can arise in LQ or $Z'$
models, scenario (I) is only possible with a $Z'$. Thus, analyses that
favor NP in $C_9^{\mu\mu}$ only are essentially favoring models in
which $\bsmumu$ arises due to $Z'$ exchange. We have performed a
model-dependent analysis of $Z'$ models, taking into account the
additional constraints from $\bs$-$\bsbar$ mixing and neutrino trident
production. If the $Z'$ is heavy, $M_{Z'} =$ O(TeV), the ${\bar\mu}\mu
Z'$ coupling is reasonably large, and could have an observable effect
in a future experiment on neutrino trident production. We also find
that a good fit to the data is found if the $Z'$ is light, $M_{Z'} =
10$ GeV or 200 MeV.

Finally, a third scenario, (III) $C_9^{\mu\mu}({\rm NP}) = -
C_{9}^{\prime \mu\mu}({\rm NP})$ has also been proposed as an explanation
for the $\bsmumu$ data. We note that this scenario predicts $R_K = 1$,
in disagreement with the experiment. In addition, this scenario can
only arise in rather contrived models. For these reasons, we exclude
scenario (III) as an explanation of the $B$-decay anomalies.

\bigskip
\noindent
{\bf Acknowledgements}: This work was financially supported by by the
U. S. Department of Energy under contract DE-SC0007983 (BB), by the
National Science Foundation under Grant No.\ PHY-1414345 (AD), and by
NSERC of Canada (DL). AD thanks Xerxes Tata for helpful conversations.
JK wishes to thank Bibhuprasad Mahakud for discussions and technical
help regarding the global fits.


\bigskip

\appendix

{\noindent\normalfont\bfseries\Large Appendix}

This Appendix contains Tables of all $\bsmumu$ experimental data used
in the fits.

\begin{table}[htb]
\centering
\begin{tabular}{|c|c|}
\hline
\multicolumn{2}{|c|}{$B^0\to K^{*0}\mu^+\mu^-$ differential branching ratio}\\
 \hline
Bin (GeV$^2$) & Measurement ($\times 10^{7}$)\\
 \hline
\multicolumn{2}{|c|}{LHCb 2016 \cite{Aaij:2016flj}}\\
 \hline
    $[1.1,2.5]$        & $0.326_{\,-0.031}^{\,+0.032} \pm {0.010} \pm0.022$\\
    $[2.5,4.0]$        & $0.334_{\,-0.033}^{\,+0.031} \pm {0.009}  \pm0.023$\\
    $[4.0,6.0]$        & $0.354_{\,-0.026}^{\,+0.027}   \pm {0.009} \pm0.024$\\
    $[15.0,19.0]$    &  $0.436_{\,-0.019}^{\,+0.018}  \pm{0.007} \pm0.030$\\
\hline
\multicolumn{2}{|c|}{CDF \cite{CDFupdate}}\\
\hline
$[0.0,2.0]$        & $0.912 \pm 1.73 \pm 0.49$\\
$[2.0,4.3]$        & $0.461\pm 1.19  \pm0.27$\\
\hline
\multicolumn{2}{|c|}{CMS 2013 \cite{Chatrchyan:2013cda}}\\
\hline
$[1.0,2.0]$        & $0.48^{+0.14}_{-0.12} \pm 0.04$\\
$[2.0,4.3]$        & $0.38\pm 0.07  \pm0.03$\\
\hline
\multicolumn{2}{|c|}{CMS 2015 \cite{Khachatryan:2015isa}}\\
\hline
$[1.0,2.0]$        & $0.46 \pm 0.07 \pm 0.03$\\
$[2.0,4.3]$        & $0.33\pm 0.05  \pm0.02$\\
\hline
\end{tabular}
\caption{Experimental measurements of the differential branching ratio
  of $B^0 \to K^{*0} \mu^+\mu^-$.}
\label{B0K*mumuBRmeas}
\end{table}

\begin{table}[htb]
\centering
\begin{tabular}{|c|c|c|}
\hline
\multicolumn{3}{|c|}{$B^0\to K^{*0}\mu^+\mu^-$ angular observables}\\
\hline
\multicolumn{3}{|c|}{ATLAS 2017 \cite{BK*mumuATLAS}}\\
\hline
$q^2 \in [\,0.04\,,\,2.0\,]\,{\rm GeV}^2 $&$q^2 \in [\,2.0\,,\,4.0\,]\,{\rm GeV}^2 $&$q^2 \in [\,4.0\,,\,6.0\,]\,{\rm GeV}^2 $ \\
\hline
$\av{F_L}=\phantom{-}0.44\, \pm 0.08 \pm 0.07$ &$\av{F_L}= \phantom{-}0.64 \pm 0.11 \pm 0.05$ &$\av{F_L}=\phantom{-}0.42 \pm 0.13 \pm 0.12$ \\
$\av{S_3}=-\phantom{-}0.02\, \pm 0.09 \pm 0.02$ &$\av{S_3}= -\phantom{-}0.15 \pm 0.10 \pm 0.07$ &$\av{S_3}=\phantom{-}0.00 \pm 0.12 \pm 0.07$ \\
$\av{S_4}=\phantom{-}0.19\, \pm 0.25 \pm 0.10$ &$\av{S_4}= -\phantom{-}0.47 \pm 0.19 \pm 0.10$ &$\av{S_4}=\phantom{-}0.40 \pm 0.21 \pm 0.09$ \\
$\av{S_5}=\phantom{-}0.33\, \pm 0.13 \pm 0.06$ &$\av{S_5}= -\phantom{-}0.16 \pm 0.15 \pm 0.05$ &$\av{S_5}=\phantom{-}0.13 \pm 0.18 \pm 0.07$ \\
$\av{S_7}=-\phantom{-}0.09\, \pm 0.10 \pm 0.02$ &$\av{S_7}= \phantom{-}0.15 \pm 0.14 \pm 0.09$ &$\av{S_7}=\phantom{-}0.03 \pm 0.13 \pm 0.07$ \\
$\av{S_8}=-\phantom{-}0.11\, \pm 0.19 \pm 0.07$ &$\av{S_8}= \phantom{-}0.41\pm 0.16 \pm 0.15$ &$\av{S_8}=-\phantom{-}0.09 \pm 0.16 \pm 0.04$ \\
\hline
\multicolumn{3}{|c|}{CMS 2017 \cite{BK*mumuCMS}}\\
\hline
$q^2 \in [\,1.0\,,\,2.0\,]\,{\rm GeV}^2 $&$q^2 \in [\,2.0\,,\,4.3\,]\,{\rm GeV}^2 $&$q^2 \in [\,4.3\,,\,6.0\,]\,{\rm GeV}^2 $ \\
\hline
$\av{P_1}=\phantom{-}0.12 \,{}^{+0.46}_{-0.47}\pm 0.06$ &$\av{P_1}= -\phantom{-}0.69 \,{}^{+0.58}_{-0.27}\pm 0.09$ &$\av{P_1}=\phantom{-}0.53 \,{}^{+0.24}_{-0.33}\pm 0.18$\\
$\av{P'_5}=\phantom{-}0.10 \,{}^{+0.32}_{-0.31}\pm 0.12$ &$\av{P'_5}= -\phantom{-}0.57 \,{}^{+0.34}_{-0.31}\pm 0.15$ &$\av{P'_5}=-\phantom{-}0.96 \,{}^{+0.22}_{-0.21}\pm 0.16$\\
\hline
\multicolumn{3}{|c|}{CMS 2015 \cite{Khachatryan:2015isa}}\\
\hline
$q^2 \in [\,1.0\,,\,2.0\,]\,{\rm GeV}^2 $&$q^2 \in [\,2.0\,,\,4.3\,]\,{\rm GeV}^2 $&$q^2 \in [\,4.3\,,\,6.0\,]\,{\rm GeV}^2 $ \\
\hline
$\av{F_L}=\phantom{-}0.64 \,{}^{+0.10}_{-0.09}\pm 0.07$ &$\av{F_L}= \phantom{-}0.80 \pm 0.08\pm 0.06$ &$\av{F_L}=\phantom{-}0.62 \,{}^{+0.10}_{-0.09}\pm 0.07$\\
$\av{A_{FB}}=-\phantom{-}0.27 \,{}^{+0.17}_{-0.40}\pm 0.07$ &$\av{A_{FB}}= -\phantom{-}0.12 \,{}^{+0.15}_{-0.17}\pm 0.05$ &$\av{A_{FB}}=-\phantom{-}0.01 \pm 0.15\pm 0.03$\\
\hline
\multicolumn{3}{|c|}{LHCb 2015 \cite{BK*mumuLHCb2}}\\
\hline
%
 $q^2 \in [\,1.1\,,\,2.5\,]\,{\rm GeV}^2 $   &$q^2 \in [\,2.5\,,\,4.0\,]\,{\rm GeV}^2 $  &$q^2 \in [\,4.0\,,\,6.0\,]\,{\rm GeV}^2 $ \\
\hline
  $\av{F_L}=\phantom{-}0.660\,{}^{+0.083}_{-0.077} \pm 0.022$&$\av{F_L}=\phantom{-}0.876\,{}^{+0.109}_{-0.097} \pm 0.017$&$\av{F_L}   =\phantom{-}0.611\,{}^{+0.052}_{-0.053} \pm 0.017$    \\
  $\av{A_{FB}}=-0.191\,{}^{+0.068}_{-0.080} \pm 0.012$ &$\av{A_{FB}}=-0.118\,{}^{+0.082}_{-0.090} \pm 0.007$   &$\av{A_{FB}}= \phantom{-}0.025\,{}^{+0.051}_{-0.052} \pm 0.004$           \\
 $\av{S_3}=-0.077\,{}^{+0.087}_{-0.105} \pm 0.005$ &$\av{S_3}=\phantom{-}0.035\,{}^{+0.098}_{-0.089} \pm 0.007$ &$\av{S_3}= \phantom{-}0.035\,{}^{+0.069}_{-0.068} \pm 0.007$         \\
 $\av{S_4}=-0.077\,{}^{+0.111}_{-0.113} \pm 0.005$    &$\av{S_4}=-0.234\,{}^{+0.127}_{-0.144} \pm 0.006$    &$\av{S_4}=-0.219\,{}^{+0.086}_{-0.084} \pm 0.008$         \\
  $\av{S_5}=\phantom{-}0.137\,{}^{+0.099}_{-0.094} \pm 0.009$ & $\av{S_5}=-0.022\,{}^{+0.110}_{-0.103} \pm 0.008$ &$\av{S_5}=-0.146\,{}^{+0.077}_{-0.078} \pm 0.011$ \\
   $\av{S_7}=-0.219\,{}^{+0.094}_{-0.104} \pm 0.004$ & $\av{S_7}=\phantom{-}0.068\,{}^{+0.120}_{-0.112} \pm 0.005$ &$\av{S_7}=-0.016\,{}^{+0.081}_{-0.080} \pm 0.004$         \\
$\av{S_8}=-0.098\,{}^{+0.108}_{-0.123} \pm 0.005$ & $\av{S_8}=\phantom{-}0.030\,{}^{+0.129}_{-0.131} \pm 0.006$ &$\av{S_8}=\phantom{-}0.167\,{}^{+0.094}_{-0.091} \pm 0.004$       \\
  $\av{S_9}=-0.119\,{}^{+0.087}_{-0.104} \pm 0.005$  & $\av{S_9}=-0.092\,{}^{+0.105}_{-0.125} \pm 0.007$ &$\av{S_9}=-0.032\,{}^{+0.071}_{-0.071} \pm 0.004$         \\
\hline
 $q^2 \in [\,15.0\,,\,19.0\,]\,{\rm GeV}^2 $&& \\
\hline
 $\av{F_L}=\phantom{-}0.344\,{}^{+0.028}_{-0.030} \pm 0.008$  &&  \\
 $\av{A_{FB}}=-\phantom{-}0.355\,{}^{+0.027}_{-0.027} \pm 0.009$ &&  \\
 $\av{S_3}=-0.163\,{}^{+0.033}_{-0.033} \pm 0.009$    &&\\
$\av{S_4}=-0.284\,{}^{+0.038}_{-0.041} \pm 0.007$         &&\\
  $\av{S_5}=-0.325\,{}^{+0.036}_{-0.037} \pm 0.009$       &&\\
 $\av{S_7}= \phantom{-}0.048\,{}^{+0.043}_{-0.043} \pm 0.006$ &&  \\
 $\av{S_8}=\phantom{-}0.028\,{}^{+0.044}_{-0.045} \pm 0.003$  &&\\
   $\av{S_9}=-0.053\,{}^{+0.039}_{-0.039} \pm 0.002$       &&\\
\hline
\multicolumn{3}{|c|}{CDF }\\
\hline
$q^2 \in [\,0.0\,,\,2.0\,]\,{\rm GeV}^2 $&$q^2 \in [\,2.0\,,\,4.3\,]\,{\rm GeV}^2 $& \\
\hline
$\av{F_L}=\phantom{-}0.26 \,{}^{+0.14}_{-0.13}\pm 0.04$ &$\av{F_L}=\phantom{-}0.72 \,{}^{+0.15}_{-0.17}\pm 0.09$ &\\
$\av{A_{FB}}=\phantom{-}0.07 \,{}^{+0.29}_{-0.28}\pm 0.11$ &$\av{A_{FB}}= -\phantom{-}0.11 \,{}^{+0.34}_{-0.45}\pm 0.16$ &\\
\hline
\hline
\end{tabular}
\caption{Experimental measurements of the angular observables of
  $B^0\to K^{*0} \mu^+\mu^-$.  }
\label{tab:BtoKstar}
\end{table}

\begin{table}[htb]
\centering
\begin{tabular}{|c|c|}
\hline
\multicolumn{2}{|c|}{$B^+\to K^{*+}\mu^+\mu^-$ differential branching ratio }\\
\hline
\multicolumn{2}{|c|}{ LHCb 2014 \cite{Aaij:2014pli}}\\
\hline
Bin (GeV$^2$) & Measurement($\times 10^{9}$)\\
\hline
$[0.1-2.0]$    & $59.2 ^{+14.4}_{-13.0} \pm 4.0$   \\
$[2.0-4.0]$    & $55.9 ^{+15.9}_{-14.4} \pm 3.8 $   \\
$[4.0-6.0]$    & $24.9 ^{+11.0}_{-9.6}  \pm 1.7 $ \\
$[15.0-19.0]$   & $39.5 ^{+8.0}_{-7.3} \pm 2.8 $\\
 \hline
\multicolumn{2}{|c|} {CDF \cite{CDFupdate}}\\
\hline
$[0.0-2.0]$    & $75.0\pm+46.8 \pm 8.8$   \\
$[2.0-4.0]$    & $49.4 \pm 35.8 \pm 6.3 $   \\
\hline
\end{tabular}
\caption{Experimental measurements of the differential branching ratio
  of $B^+ \to K^{*+} \mu^+\mu^-$.}
\label{B+K*mumuBRmeas}
\end{table}

\begin{table}[htb]
\centering
\begin{tabular}{|c|c|}
\hline
\multicolumn{2}{|c|}{$B^+\to K^{+}\mu^+\mu^-$ differential branching ratio}\\
 \hline
\multicolumn{2}{|c|}{LHCb 2014 \cite{Aaij:2014pli}}\\
 \hline
Bin (GeV$^2$) & Measurement ($\times 10^{9}$)\\
\hline
$[1.1-2.0]$    & $23.3 \pm 1.5 \pm 1.2$ \\
$[2.0-3.0]$     & $28.2 \pm 1.6 \pm 1.4$ \\
$[3.0-4.0]$    & $25.4 \pm 1.5 \pm 1.3$ \\
$[4.0-5.0]$   & $22.1 \pm 1.4 \pm 1.1$ \\
$[5.0-6.0]$     & $23.1 \pm 1.4 \pm 1.2$ \\
$[15.0-22.0]$   & $\phantom{0}12.1 \pm 0.4 \pm 0.6$  \\
\hline
\multicolumn{2}{|c|}{CDF \cite{CDFupdate}}\\
\hline
$[0.0-2.0]$    & $18.0 \pm 5.3 \pm 1.2$ \\
$[2.0-4.3]$     & $31.6 \pm 5.4 \pm 1.8$ \\
\hline
\end{tabular}
\caption{Experimental measurements of the differential branching
  ratio of $B^+\to K^{+} \mu^+\mu^-$ . }
\label{B+KmumuBRmeas}
\end{table}

\begin{table}
\centering
\begin{tabular}{|c|c|}
\hline
\multicolumn{2}{|c|}{$B^0\to K^{0}\mu^+\mu^-$ differential branching ratio }\\
 \hline
\multicolumn{2}{|c|}{LHCb 2014 \cite{Aaij:2014pli}}\\
 \hline
Bin (GeV$^2$) & Measurement ($\times 10^{9}$)\\
\hline
$[0.1-2.0]$    & $12.2 ^{+5.9}_{-5.2} \pm 0.6$ \\
$[2.0-4.0]$    & $18.7 ^{+5.5}_{-4.9} \pm 0.9 $ \\
$[4.0-6.0]$  & $17.3 ^{+5.3}_{-4.8} \pm 0.9 $ \\
$[15.0-22.0]$  & $ \phantom{0}9.5 ^{+1.6}_{-1.5} \pm 0.5 $ \\
\hline
\multicolumn{2}{|c|}{CDF \cite{CDFupdate}}\\
\hline
$[0.0-2.0]$    & $24.5 \pm 15.9 \pm 2.1$ \\
$[2.0-4.3]$    & $25.5 \pm 17.0 \pm 3.5 $ \\
\hline
\end{tabular}
\caption{Experimental measurements of the differential branching
  ratio of $B^0\to K^{0} \mu^+\mu^-$. }
\label{B0KmumuBRmeas}
\end{table}

\begin{table}
\centering
\begin{tabular}{|c|c|}
\hline
\multicolumn{2}{|c|}{$\bs \to \phi \mu^+\mu^-$ differential branching ratio}\\
 \hline
Bin (GeV$^2$) & Measurement ($\times 10^{8}$)\\
\hline
$[1.0-6.0]$   & $2.58 ^{+0.33}_{-0.31} \pm{0.08} \pm{0.19}$ \\
$[15.0-19.0]$ & $4.04 ^{+0.39}_{-0.38} \pm{0.13} \pm{0.30}$ \\
\hline
\end{tabular}
\caption{Experimental measurements of the differential branching ratio
  of $\bs \to \phi \mu^+\mu^-$ \cite{BsphimumuLHCb2}. The experimental
  errors are, from left to right, statistical, systematic and due to
  the uncertainty on the branching ratio of the normalization mode
  $\bs \to J/\psi\phi$.}
\label{BsphimumuBRmeas}
\end{table}

\begin{table}[t]
\centering
\begin{tabular}{|c|c|}
\hline
\multicolumn{2}{|c|}{$\bs \to \phi \mu^+\mu^-$ angular observables}\\
\hline
%
$q^2 \in [\,0.1\,,\,2.0\,]\,{\rm GeV}^2 $  & $q^2 \in [\,2.0\,,\,5.0\,]\,{\rm GeV}^2 $        \\
\hline
$\av{F_L}   = \phantom{-} 0.20 ^{+0.08}_{-0.09}\pm 0.02 $      & $\av{F_L}  = \phantom{-} 0.68 ^{+ 0.16 }_{ -0.13 } \pm 0.03 $ \\
$\av{S_3}    =  -0.05 ^{+ 0.13 }_{ -0.13 } \pm 0.01 $           &  $\av{S_3}=  -0.06 ^{+ 0.19 }_{ -0.23 } \pm 0.01 $        \\
$\av{S_4}    = \phantom{-} 0.27 ^{+ 0.28 }_{ -0.18 } \pm 0.01 $ &  $\av{S_4}= -0.47 ^{+ 0.30 }_{ -0.44 } \pm 0.01 $ \\
$\av{S_7}   = \phantom{-} 0.04^{+ 0.12 }_{ -0.12 } \pm 0.00 $  &  $\av{S_7}= -0.03^{+ 0.18 }_{ -0.23 } \pm 0.01 $      \\
\hline
$q^2 \in [\,15.0\,,\,19.0\,]\,{\rm GeV}^2 $  &    \\
\hline
$\av{F_L}=\phantom{-} 0.29 ^{+ 0.07 }_{ -0.06 } \pm 0.02 $&    \\
$\av{S_3}=-0.09 ^{+ 0.11 }_{ -0.12 } \pm 0.01 $      &          \\
$\av{S_4}= -0.14 ^{+ 0.11 }_{ -0.11 } \pm 0.01 $     & \\
$\av{S_7}= \phantom{-} 0. 13^{+ 0.11 }_{ -0.11 } \pm 0.01 $  & \\
\hline
\end{tabular}
\caption{Experimental measurements of the angular observables of
  $\bs \to \phi \mu^+\mu^-$ \cite{BsphimumuLHCb2}.  The experimental
  errors are, from left to right, statistical and systematic.}
\label{Bsphimumuangmeas}
\end{table}

\begin{table}
\centering
\begin{tabular}{|c|c|}
\hline
\multicolumn{2}{|c|}{$B\to X_s \mu^+ \mu^-$ differential branching ratio}\\
 \hline
Bin & Measurement ($\times 10^{6}$)\\
\hline
$q^2 \in [1,6] ~{\rm GeV}^2$    & $0.66 \pm{0.88}$ \\
$q^2 > 14.2 ~{\rm GeV}^2$  & $0.60 \pm{0.31}$ \\
\hline
\end{tabular}
\caption{Experimental measurements of the differential branching ratio
  of $B\to X_s \mu^+ \mu^-$ \cite{Lees:2013nxa}.}
\label{BXsmumuBRmeas}
\end{table}

\pagebreak


\clearpage

\begin{thebibliography}{99}

\bibitem{RK*expt} S. Bifani (on behalf of the LHCb Collaboration),
``Search for New Physics with $\bsll$ decays at LHCb,''
talk given at CERN, April 18, 2017.

\bibitem{RK*theory} See, for example,
G.~Hiller and F.~Kruger,
  ``More model-independent analysis of $b \to s$ processes,''
  Phys.\ Rev.\ D {\bf 69}, 074020 (2004)
  doi:10.1103/PhysRevD.69.074020
  [hep-ph/0310219].

\bibitem{flavio}
David Straub, \textit{flavio v0.11, 2016.}
  \href{http://dx.doi.org/10.5281/zenodo.59840}{http://dx.doi.org/10.5281/zenodo.59840}

\bibitem{RKexpt} R.~Aaij {\it et al.}  [LHCb Collaboration],
  ``Test of lepton universality using $B^{+}\rightarrow K^{+}\ell^{+}\ell^{-}$ decays,''
  Phys.\ Rev.\ Lett.\  {\bf 113}, 151601 (2014)
  [arXiv:1406.6482 [hep-ex]].

\bibitem{IsidoriRK}
M.~Bordone, G.~Isidori and A.~Pattori,
  ``On the Standard Model predictions for $R_K$ and $R_{K^*}$,''
  Eur.\ Phys.\ J.\ C {\bf 76}, no. 8, 440 (2016)
  doi:10.1140/epjc/s10052-016-4274-7
  [arXiv:1605.07633 [hep-ph]].

\bibitem{bslltheorerror} See, for example,
V.~G.~Chobanova, T.~Hurth, F.~Mahmoudi, D.~Martinez Santos and S.~Neshatpour,
  ``Large hadronic power corrections or new physics in the rare decay $B \to K^* \ell\ell$?,''
  arXiv:1702.02234 [hep-ph], and references therein.

\bibitem{BK*mumuLHCb1}
R.~Aaij {\it et al.} [LHCb Collaboration],
  ``Measurement of Form-Factor-Independent Observables in the Decay $B^{0} \to K^{*0} \mu^+ \mu^-$,''
  Phys.\ Rev.\ Lett.\  {\bf 111}, 191801 (2013)
  doi:10.1103/PhysRevLett.111.191801
  [arXiv:1308.1707 [hep-ex]].

\bibitem{BK*mumuLHCb2}
R.~Aaij {\it et al.} [LHCb Collaboration],
  ``Angular analysis of the $B^{0} \to K^{*0} \mu^{+} \mu^{-}$ decay using 3 fb$^{-1}$ of integrated luminosity,''
  JHEP {\bf 1602}, 104 (2016)
  doi:10.1007/JHEP02(2016)104
  [arXiv:1512.04442 [hep-ex]].

\bibitem{BK*mumuBelle}
A.~Abdesselam {\it et al.} [Belle Collaboration],
  ``Angular analysis of $B^0 \to K^\ast(892)^0 \ell^+ \ell^-$,''
  arXiv:1604.04042 [hep-ex].

\bibitem{P'5} S.~Descotes-Genon, T.~Hurth, J.~Matias and J.~Virto,
  ``Optimizing the basis of $B \to K^* l l$ observables in the full kinematic range,''
  JHEP {\bf 1305}, 137 (2013)
  doi:10.1007/JHEP05(2013)137
  [arXiv:1303.5794 [hep-ph]].

\bibitem{BK*mumuATLAS}
ATLAS Collaboration,
``Angular analysis of $B_d^0 \to K^* \mu^+ \mu^-$ decays in $pp$ collisions at $\sqrt{s} = 8$ TeV with the ATLAS detector,''
Tech.\ Rep.\ ATLAS-CONF-2017-023, CERN, Geneva, 2017.

\bibitem{BK*mumuCMS}
CMS Collaboration,
``Measurement of the $P_1$ and $P'_5$ angular parameters of the decay $B^0 \to K^{*0} \mu^+ \mu^-$
in proton-proton collisions at $\sqrt{s} = 8$ TeV,''
Tech.\ Rep.\ CMS-PAS-BPH-15-008, CERN, Geneva, 2017.

\bibitem{BsphimumuLHCb1}
R.~Aaij {\it et al.} [LHCb Collaboration],
  ``Differential branching fraction and angular analysis of the decay $B_s^0\to\phi\mu^{+}\mu^{-}$,''
  JHEP {\bf 1307}, 084 (2013)
  doi:10.1007/JHEP07(2013)084
  [arXiv:1305.2168 [hep-ex]].

\bibitem{BsphimumuLHCb2}
R.~Aaij {\it et al.} [LHCb Collaboration],
  ``Angular analysis and differential branching fraction of the decay $B^0_s\to\phi\mu^+\mu^-$,''
  JHEP {\bf 1509}, 179 (2015)
  doi:10.1007/JHEP09(2015)179
  [arXiv:1506.08777 [hep-ex]].

\bibitem{latticeQCD1}
R.~R.~Horgan, Z.~Liu, S.~Meinel and M.~Wingate,
  ``Calculation of $B^0 \to K^{*0} \mu^+ \mu^-$ and $B_s^0 \to \phi \mu^+ \mu^-$ observables using form factors from lattice QCD,''
  Phys.\ Rev.\ Lett.\  {\bf 112}, 212003 (2014)
  doi:10.1103/PhysRevLett.112.212003
  [arXiv:1310.3887 [hep-ph]],

\bibitem{latticeQCD2}
  ``Rare $B$ decays using lattice QCD form factors,''
  PoS LATTICE {\bf 2014}, 372 (2015)
  [arXiv:1501.00367 [hep-lat]].

\bibitem{QCDsumrules}
A.~Bharucha, D.~M.~Straub and R.~Zwicky,
  ``$B\to V\ell^+\ell^-$ in the Standard Model from light-cone sum rules,''
  JHEP {\bf 1608}, 098 (2016)
  doi:10.1007/JHEP08(2016)098
  [arXiv:1503.05534 [hep-ph]].

\bibitem{Capdevila:2017bsm}
  B.~Capdevila, A.~Crivellin, S.~Descotes-Genon, J.~Matias and J.~Virto,
  ``Patterns of New Physics in $b\to s\ell^+\ell^-$ transitions in the light of recent data,''
  arXiv:1704.05340 [hep-ph].

\bibitem{Altmannshofer:2017yso}
  W.~Altmannshofer, P.~Stangl and D.~M.~Straub,
  ``Interpreting Hints for Lepton Flavor Universality Violation,''
  arXiv:1704.05435 [hep-ph].

\bibitem{DAmico:2017mtc}
  G.~D'Amico, M.~Nardecchia, P.~Panci, F.~Sannino, A.~Strumia, R.~Torre and A.~Urbano,
  ``Flavour anomalies after the $R_{K^*}$ measurement,''
  arXiv:1704.05438 [hep-ph].

\bibitem{Hiller:2017bzc}
  G.~Hiller and I.~Nisandzic,
  ``$R_K$ and $R_{K^{\ast}}$ beyond the Standard Model,''
  arXiv:1704.05444 [hep-ph].

\bibitem{Geng:2017svp}
  L.~S.~Geng, B.~Grinstein, S.~Jäger, J.~Martin Camalich, X.~L.~Ren and R.~X.~Shi,
  ``Towards the discovery of new physics with lepton-universality ratios of $b\to s\ell\ell$ decays,''
  arXiv:1704.05446 [hep-ph].

\bibitem{Ciuchini:2017mik}
  M.~Ciuchini, A.~M.~Coutinho, M.~Fedele, E.~Franco, A.~Paul, L.~Silvestrini and M.~Valli,
  ``On Flavourful Easter eggs for New Physics hunger and Lepton Flavour Universality violation,''
  arXiv:1704.05447 [hep-ph].

\bibitem{Celis:2017doq}
  A.~Celis, J.~Fuentes-Martin, A.~Vicente and J.~Virto,
  ``Gauge-invariant implications of the LHCb measurements on Lepton-Flavour Non-Universality,''
  arXiv:1704.05672 [hep-ph].

\bibitem{DiChiara:2017cjq}
  S.~Di Chiara, A.~Fowlie, S.~Fraser, C.~Marzo, L.~Marzola, M.~Raidal and C.~Spethmann,
  ``Minimal flavor-changing $Z'$ models and muon $g-2$ after the $R_{K^*}$ measurement,''
  arXiv:1704.06200 [hep-ph].

\bibitem{Sala:2017ihs}
  F.~Sala and D.~M.~Straub,
  ``A New Light Particle in B Decays?,''
  arXiv:1704.06188 [hep-ph].

\bibitem{Ghosh:2017ber}
  D.~Ghosh,
  ``Explaining the $R_K$ and $R_{K^*}$ anomalies,''
  arXiv:1704.06240 [hep-ph].

\bibitem{BK*mumuhadunc1}
S.~Descotes-Genon, L.~Hofer, J.~Matias and J.~Virto,
  ``On the impact of power corrections in the prediction of $B \to K^*\mu^+\mu^-$ observables,''
  JHEP {\bf 1412}, 125 (2014)
  doi:10.1007/JHEP12(2014)125
  [arXiv:1407.8526 [hep-ph]].

\bibitem{BK*mumuhadunc2}
J.~Lyon and R.~Zwicky,
  ``Resonances gone topsy turvy - the charm of QCD or new physics in $b \to s \ell^+ \ell^-$?,''
  arXiv:1406.0566 [hep-ph].

\bibitem{BK*mumuhadunc3}
S.~J{\"a}ger and J.~Martin Camalich,
  ``Reassessing the discovery potential of the $B \to K^{*} \ell^+\ell^-$ decays in the large-recoil region: SM challenges and BSM opportunities,''
  Phys.\ Rev.\ D {\bf 93}, 014028 (2016)
  doi:10.1103/PhysRevD.93.014028
  [arXiv:1412.3183 [hep-ph]].

\bibitem{bsmumuNPCPC}
A.~K.~Alok, A.~Datta, A.~Dighe, M.~Duraisamy, D.~Ghosh and D.~London,
  ``New Physics in $b \to s \mu^+ \mu^-$: CP-Conserving Observables,''
  JHEP {\bf 1111}, 121 (2011)
  doi:10.1007/JHEP11(2011)121
  [arXiv:1008.2367 [hep-ph]].

\bibitem{bsmumuNPCPV}
A.~K.~Alok, A.~Datta, A.~Dighe, M.~Duraisamy, D.~Ghosh and D.~London,
  ``New Physics in $b \to s \mu^+ \mu^-$: CP-Violating Observables,''
  JHEP {\bf 1111}, 122 (2011)
  doi:10.1007/JHEP11(2011)122
  [arXiv:1103.5344 [hep-ph]].

\bibitem{Bardhan:2017xcc} 
  D.~Bardhan, P.~Byakti and D.~Ghosh,
  ``Role of Tensor operators in $R_K$ and $R_{K^*}$,''
  Phys.\ Lett.\ B {\bf 773}, 505 (2017)
  doi:10.1016/j.physletb.2017.08.062
  [arXiv:1705.09305 [hep-ph]].

\bibitem{Datta:2013kja}
  A.~Datta, M.~Duraisamy and D.~Ghosh,
  ``Explaining the $B \to K^\ast \mu^+ \mu^-$ data with scalar interactions,''
  Phys.\ Rev.\ D {\bf 89}, no. 7, 071501 (2014)
  doi:10.1103/PhysRevD.89.071501
  [arXiv:1310.1937 [hep-ph]].

\bibitem{CCO}
L.~Calibbi, A.~Crivellin and T.~Ota,
  ``Effective Field Theory Approach to $b\to s \ell\ell^{(\prime)}$, $B\to K^{(*)} \nu {\bar\nu}$ and $B \to D^{(*)} \tau\nu$ with Third Generation Couplings,''
  Phys.\ Rev.\ Lett.\  {\bf 115}, 181801 (2015)
  doi:10.1103/PhysRevLett.115.181801
  [arXiv:1506.02661 [hep-ph]].

\bibitem{AGC}
R.~Alonso, B.~Grinstein and J.~Martin Camalich,
  ``Lepton universality violation and lepton flavor conservation in $B$-meson decays,''
  JHEP {\bf 1510}, 184 (2015)
  doi:10.1007/JHEP10(2015)184
  [arXiv:1505.05164 [hep-ph]].

\bibitem{HS1}
  G.~Hiller and M.~Schmaltz,
  ``$R_K$ and future $b \to s \ell \ell$ BSM opportunities,''
  Phys.\ Rev.\ D {\bf 90} (2014) 054014
  [arXiv:1408.1627 [hep-ph]].

\bibitem{GNR}
B.~Gripaios, M.~Nardecchia and S.~A.~Renner,
  ``Composite leptoquarks and anomalies in $B$-meson decays,''
  JHEP {\bf 1505}, 006 (2015)
  doi:10.1007/JHEP05(2015)006
  [arXiv:1412.1791 [hep-ph]].

\bibitem{VH}
I.~de Medeiros Varzielas and G.~Hiller,
  ``Clues for flavor from rare lepton and quark decays,''
  JHEP {\bf 1506}, 072 (2015)
  doi:10.1007/JHEP06(2015)072
  [arXiv:1503.01084 [hep-ph]].

\bibitem{SM}
S.~Sahoo and R.~Mohanta,
  ``Scalar leptoquarks and the rare $B$ meson decays,''
  Phys.\ Rev.\ D {\bf 91}, no. 9, 094019 (2015)
  doi:10.1103/PhysRevD.91.094019
  [arXiv:1501.05193 [hep-ph]].

\bibitem{FK}
S.~Fajfer and N.~Ko{\v s}nik,
  ``Vector leptoquark resolution of $R_K$ and $R_{D^{(*)}}$ puzzles,''
  Phys.\ Lett.\ B {\bf 755}, 270 (2016)
  doi:10.1016/j.physletb.2016.02.018
  [arXiv:1511.06024 [hep-ph]].

\bibitem{BFK}
D.~Be{\v c}irevi{\' c}, S.~Fajfer and N.~Ko{\v s}nik,
  ``Lepton flavor nonuniversality in $b \to s \ell^+\ell^-$ processes,''
  Phys.\ Rev.\ D {\bf 92}, no. 1, 014016 (2015)
  doi:10.1103/PhysRevD.92.014016
  [arXiv:1503.09024 [hep-ph]].

\bibitem{BKSZ}
D.~Be{\v c}irevi{\' c}, N.~Ko{\v s}nik, O.~Sumensari and R.~Zukanovich Funchal,
  ``Palatable Leptoquark Scenarios for Lepton Flavor Violation in Exclusive $b\to s\ell_1\ell_2$ modes,''
  JHEP {\bf 1611}, 035 (2016)
  doi:10.1007/JHEP11(2016)035
  [arXiv:1608.07583 [hep-ph]].

 \bibitem{Crivellin:2015lwa}
  A.~Crivellin, G.~D'Ambrosio and J.~Heeck,
  ``Addressing the LHC flavor anomalies with horizontal gauge symmetries,''
  Phys.\ Rev.\ D {\bf 91}, 075006 (2015)
  doi:10.1103/PhysRevD.91.075006
  [arXiv:1503.03477 [hep-ph]].

\bibitem{Isidori} A.~Greljo, G.~Isidori and D.~Marzocca,
  ``On the breaking of Lepton Flavor Universality in B decays,''
  JHEP {\bf 1507}, 142 (2015)
  doi:10.1007/JHEP07(2015)142
  [arXiv:1506.01705 [hep-ph]].

  \bibitem{dark}
  D.~Aristizabal Sierra, F.~Staub and A.~Vicente,
  ``Shedding light on the $b\to s$ anomalies with a dark sector,''
  Phys.\ Rev.\ D {\bf 92}, 015001 (2015)
  doi:10.1103/PhysRevD.92.015001
  [arXiv:1503.06077 [hep-ph]].

\bibitem{Chiang}
C.~W.~Chiang, X.~G.~He and G.~Valencia,
``$Z'$ model for $b \to s \ell {\bar\ell}$ flavor anomalies,''
  Phys.\ Rev.\ D {\bf 93}, 074003 (2016)
  doi:10.1103/PhysRevD.93.074003
  [arXiv:1601.07328 [hep-ph]].

\bibitem{Virto}
S.~M.~Boucenna, A.~Celis, J.~Fuentes-Martin, A.~Vicente and J.~Virto,
  ``Non-abelian gauge extensions for $B$-decay anomalies,''
  Phys.\ Lett.\ B {\bf 760}, 214 (2016)
  doi:10.1016/j.physletb.2016.06.067
  [arXiv:1604.03088 [hep-ph]],
  ``Phenomenology of an $SU(2) \times SU(2) \times U(1)$ model with lepton-flavour non-universality,''
  JHEP {\bf 1612}, 059 (2016)
  doi:10.1007/JHEP12(2016)059
  [arXiv:1608.01349 [hep-ph]].

\bibitem{GGH}
R.~Gauld, F.~Goertz and U.~Haisch,
  ``On minimal $Z'$ explanations of the $B\to K^*\mu^+\mu^-$ anomaly,''
  Phys.\ Rev.\ D {\bf 89}, 015005 (2014)
  doi:10.1103/PhysRevD.89.015005
  [arXiv:1308.1959 [hep-ph]],
  ``An explicit $Z'$-boson explanation of the $B \to K^* \mu^+ \mu^-$ anomaly,''
  JHEP {\bf 1401}, 069 (2014)
  doi:10.1007/JHEP01(2014)069
  [arXiv:1310.1082 [hep-ph]].

\bibitem{BG}
A.~J.~Buras and J.~Girrbach,
  ``Left-handed $Z'$ and $Z$ FCNC quark couplings facing new $\bsmumu$ data,''
  JHEP {\bf 1312}, 009 (2013)
  doi:10.1007/JHEP12(2013)009
  [arXiv:1309.2466 [hep-ph]].

\bibitem{BFG}
A.~J.~Buras, F.~De Fazio and J.~Girrbach,
  ``331 models facing new $b \to s\mu^+ \mu^-$ data,''
  JHEP {\bf 1402}, 112 (2014)
  doi:10.1007/JHEP02(2014)112
  [arXiv:1311.6729 [hep-ph]].

\bibitem{Perimeter}
W.~Altmannshofer, S.~Gori, M.~Pospelov and I.~Yavin,
  ``Quark flavor transitions in $L_\mu-L_\tau$ models,''
  Phys.\ Rev.\ D {\bf 89}, 095033 (2014)
  doi:10.1103/PhysRevD.89.095033
  [arXiv:1403.1269 [hep-ph]].

\bibitem{CDH}
A.~Crivellin, G.~D'Ambrosio and J.~Heeck,
  ``Explaining $h\to\mu^\pm\tau^\mp$, $B\to K^* \mu^+\mu^-$ and $B\to K \mu^+\mu^-/B\to K e^+e^-$ in a two-Higgs-doublet model with gauged $L_\mu-L_\tau$,''
  Phys.\ Rev.\ Lett.\  {\bf 114}, 151801 (2015)
  doi:10.1103/PhysRevLett.114.151801
  [arXiv:1501.00993 [hep-ph]],
``Addressing the LHC flavor anomalies with horizontal gauge symmetries,''
  Phys.\ Rev.\ D {\bf 91}, no. 7, 075006 (2015)
  doi:10.1103/PhysRevD.91.075006
  [arXiv:1503.03477 [hep-ph]].

\bibitem{SSV}
D.~Aristizabal Sierra, F.~Staub and A.~Vicente,
  ``Shedding light on the $b\to s$ anomalies with a dark sector,''
  Phys.\ Rev.\ D {\bf 92}, no. 1, 015001 (2015)
  doi:10.1103/PhysRevD.92.015001
  [arXiv:1503.06077 [hep-ph]].

\bibitem{CHMNPR}
A.~Crivellin, L.~Hofer, J.~Matias, U.~Nierste, S.~Pokorski and J.~Rosiek,
  ``Lepton-flavour violating $B$ decays in generic $Z'$ models,''
  Phys.\ Rev.\ D {\bf 92}, no. 5, 054013 (2015)
  doi:10.1103/PhysRevD.92.054013
  [arXiv:1504.07928 [hep-ph]].

\bibitem{CMJS}
A.~Celis, J.~Fuentes-Martin, M.~Jung and H.~Serodio,
  ``Family nonuniversal $Z'$ models with protected flavor-changing interactions,''
  Phys.\ Rev.\ D {\bf 92}, no. 1, 015007 (2015)
  doi:10.1103/PhysRevD.92.015007
  [arXiv:1505.03079 [hep-ph]].

\bibitem{BDW}
G.~B\'elanger, C.~Delaunay and S.~Westhoff,
  ``A Dark Matter Relic From Muon Anomalies,''
  Phys.\ Rev.\ D {\bf 92}, 055021 (2015)
  doi:10.1103/PhysRevD.92.055021
  [arXiv:1507.06660 [hep-ph]].

\bibitem{FNZ}
A.~Falkowski, M.~Nardecchia and R.~Ziegler,
  ``Lepton Flavor Non-Universality in $B$-meson Decays from a $U(2)$ Flavor Model,''
  JHEP {\bf 1511}, 173 (2015)
  doi:10.1007/JHEP11(2015)173
  [arXiv:1509.01249 [hep-ph]].

\bibitem{Carmona:2015ena}
  A.~Carmona and F.~Goertz,
  ``Lepton Flavor and Nonuniversality from Minimal Composite Higgs Setups,''
  Phys.\ Rev.\ Lett.\  {\bf 116}, no. 25, 251801 (2016)
  doi:10.1103/PhysRevLett.116.251801
  [arXiv:1510.07658 [hep-ph]].

\bibitem{AQSS}
B.~Allanach, F.~S.~Queiroz, A.~Strumia and S.~Sun,
  ``$Z'$ models for the LHCb and $g-2$ muon anomalies,''
  Phys.\ Rev.\ D {\bf 93}, no. 5, 055045 (2016)
  doi:10.1103/PhysRevD.93.055045
  [arXiv:1511.07447 [hep-ph]].

\bibitem{CFL}
A.~Celis, W.~Z.~Feng and D.~L\"ust,
  ``Stringy explanation of $b\to s\ell^+ \ell^-$ anomalies,''
  JHEP {\bf 1602}, 007 (2016)
  doi:10.1007/JHEP02(2016)007
  [arXiv:1512.02218 [hep-ph]].

\bibitem{Hou}
K.~Fuyuto, W.~S.~Hou and M.~Kohda,
  ``$Z'$-induced FCNC decays of top, beauty, and strange quarks,''
  Phys.\ Rev.\ D {\bf 93}, no. 5, 054021 (2016)
  doi:10.1103/PhysRevD.93.054021
  [arXiv:1512.09026 [hep-ph]].

\bibitem{CHV}
C.~W.~Chiang, X.~G.~He and G.~Valencia,
  ``$Z'$ model for $b\to s \ell {\bar\ell}$ flavor anomalies,''
  Phys.\ Rev.\ D {\bf 93}, no. 7, 074003 (2016)
  doi:10.1103/PhysRevD.93.074003
  [arXiv:1601.07328 [hep-ph]].

\bibitem{CFV}
A.~Celis, W.~Z.~Feng and M.~Vollmann,
  ``Dirac Dark Matter and $b \to s \ell^+ \ell^-$ with $\mathbf{U(1)}$ gauge symmetry,''
  arXiv:1608.03894 [hep-ph].

\bibitem{CFGI}
A.~Crivellin, J.~Fuentes-Martin, A.~Greljo and G.~Isidori,
  ``Lepton Flavor Non-Universality in $B$ decays from Dynamical Yukawas,''
  arXiv:1611.02703 [hep-ph].

\bibitem{IGG}
I.~Garcia Garcia,
  ``LHCb anomalies from a natural perspective,''
  arXiv:1611.03507 [hep-ph].

\bibitem{BdecaysDM}
J.~M.~Cline, J.~M.~Cornell, D.~London and R.~Watanabe,
  ``Hidden sector explanation of $B$-decay and cosmic ray anomalies,''
  arXiv:1702.00395 [hep-ph].

  \bibitem{ZMeV}
A.~Datta, J.~Liao and D.~Marfatia,
  ``A light $Z^\prime$ for the $R_K$ puzzle and nonstandard neutrino interactions,''
  Phys.\ Lett.\ B {\bf 768}, 265 (2017)
  doi:10.1016/j.physletb.2017.02.058
  [arXiv:1702.01099 [hep-ph]].

\bibitem{Megias:2017ove}
  E.~Megias, M.~Quiros and L.~Salas,
  ``Lepton-flavor universality violation in $R_{D^{(*)}}$ and $R_K$ from warped space,''
  arXiv:1703.06019 [hep-ph].

\bibitem{Ahmed:2017vsr}
  I.~Ahmed and A.~Rehman,
  ``LHCb anomaly in $\boldsymbol{B\to K^*\mu^+ \mu^-}$ optimised observables and potential of $\boldsymbol{Z^\prime}$ Model,''
  arXiv:1703.09627 [hep-ph].

\bibitem{RKRDmodels}
B.~Bhattacharya, A.~Datta, J.~P.~Gu\'evin, D.~London and R.~Watanabe,
  ``Simultaneous Explanation of the $R_K$ and $R_{D^{(*)}}$ Puzzles: a Model Analysis,''
  arXiv:1609.09078 [hep-ph].

\bibitem{trident}
   W.~Altmannshofer, S.~Gori, M.~Pospelov and I.~Yavin,
  ``Neutrino Trident Production: A Powerful Probe of New Physics with Neutrino Beams,''
  Phys.\ Rev.\ Lett.\  {\bf 113}, 091801 (2014)
  doi:10.1103/PhysRevLett.113.091801
  [arXiv:1406.2332 [hep-ph]].

\bibitem{bsmumuCPV}
A.~K.~Alok, B.~Bhattacharya, D.~Kumar, J.~Kumar, D.~London and S.~U.~Sankar,
  ``New Physics in $b \rightarrow s \mu^+ \mu^-$: Distinguishing Models through CP-Violating Effects,''
  arXiv:1703.09247 [hep-ph].

\bibitem{James:1975dr}
  F.~James and M.~Roos,
  ``Minuit: A System for Function Minimization and Analysis of the Parameter Errors and Correlations,''
  Comput.\ Phys.\ Commun.\  {\bf 10}, 343 (1975).
  doi:10.1016/0010-4655(75)90039-9

\bibitem{James:2004xla}
  F.~James and M.~Winkler,
  ``MINUIT User's Guide,''

\bibitem{James:1994vla}
  F.~James,
  ``MINUIT Function Minimization and Error Analysis:  Reference Manual Version 94.1,''
  CERN-D-506, CERN-D506.

\bibitem{Aaij:2013aka}
R.~Aaij {\it et al.} [LHCb Collaboration],
  ``Measurement of the $B^0_s \to \mu^+ \mu^-$ branching fraction and search for $B^0 \to \mu^+ \mu^-$ decays at the LHCb experiment,''
  Phys.\ Rev.\ Lett.\  {\bf 111}, 101805 (2013)
  doi:10.1103/PhysRevLett.111.101805
  [arXiv:1307.5024 [hep-ex]].

\bibitem{CMS:2014xfa}
V.~Khachatryan {\it et al.} [CMS and LHCb Collaborations],
  ``Observation of the rare $B^0_s\to\mu^+\mu^-$ decay from the combined analysis of CMS and LHCb data,''
  Nature {\bf 522}, 68 (2015)
  doi:10.1038/nature14474
  [arXiv:1411.4413 [hep-ex]].

\bibitem{Altmannshofer:2017fio}
  W.~Altmannshofer, C.~Niehoff, P.~Stangl and D.~M.~Straub,
  ``Status of the $B\to K^*\mu^+\mu^-$ anomaly after Moriond 2017,''
  arXiv:1703.09189 [hep-ph].

\bibitem{Beneke:2001at}
  M.~Beneke, T.~Feldmann and D.~Seidel,
  ``Systematic approach to exclusive $B \to  V l^+ l^-$, $V \gamma$ decays,''
  Nucl.\ Phys.\ B {\bf 612}, 25 (2001)
  doi:10.1016/S0550-3213(01)00366-2
  [hep-ph/0106067].

 \bibitem{Beylich:2011aq}
  M.~Beylich, G.~Buchalla and T.~Feldmann,
  ``Theory of $B \to K^{(*)}\ell^+ \ell^-$ decays at high $q^2$: OPE and quark-hadron duality,''
  Eur.\ Phys.\ J.\ C {\bf 71}, 1635 (2011)
  doi:10.1140/epjc/s10052-011-1635-0
  [arXiv:1101.5118 [hep-ph]].

\bibitem{Sakakietal}
Y.~Sakaki, M.~Tanaka, A.~Tayduganov and R.~Watanabe,
  ``Testing leptoquark models in $\bar B \to D^{(*)} \tau \bar\nu$,''
  Phys.\ Rev.\ D {\bf 88}, no. 9, 094012 (2013)
  doi:10.1103/PhysRevD.88.094012
  [arXiv:1309.0301 [hep-ph]].

\bibitem{LQmasslimits}
G.~Aad {\it et al.} [ATLAS Collaboration],
  ``Searches for scalar leptoquarks in pp collisions at ${\sqrt{s}}$ = 8 TeV with the ATLAS detector,''
  Eur.\ Phys.\ J.\ C {\bf 76}, no. 1, 5 (2016)
  doi:10.1140/epjc/s10052-015-3823-9
  [arXiv:1508.04735 [hep-ex]].

\bibitem{RKRD}
B.~Bhattacharya, A.~Datta, D.~London and S.~Shivashankara,
  ``Simultaneous Explanation of the $R_K$ and $R(D^{(*)})$ Puzzles,''
  Phys.\ Lett.\ B {\bf 742}, 370 (2015)
  [arXiv:1412.7164 [hep-ph]].

\bibitem{RD_BaBar}
  J.~P.~Lees {\it et al.} [BaBar Collaboration],
  ``Measurement of an Excess of $\bar{B} \to D^{(*)}\tau^- \bar{\nu}_\tau$ Decays and Implications for Charged Higgs Bosons,''
  Phys.\ Rev.\ D {\bf 88}, 072012 (2013)
  doi:10.1103/PhysRevD.88.072012
  [arXiv:1303.0571 [hep-ex]].

\bibitem{RD_Belle}
M.~Huschle {\it et al.} [Belle Collaboration],
  ``Measurement of the branching ratio of $\bar{B} \to D^{(\ast)} \tau^- \bar{\nu}_\tau$ relative to $\bar{B} \to D^{(\ast)} \ell^- \bar{\nu}_\ell$ decays with hadronic tagging at Belle,''
  Phys.\ Rev.\ D {\bf 92}, 072014 (2015)
  doi:10.1103/PhysRevD.92.072014
  [arXiv:1507.03233 [hep-ex]].

\bibitem{RD_LHCb}
R.~Aaij {\it et al.} [LHCb Collaboration],
  ``Measurement of the ratio of branching fractions $\mathcal{B}(\bar{B}^0 \to D^{*+}\tau^{-}\bar{\nu}_{\tau})/\mathcal{B}(\bar{B}^0 \to D^{*+}\mu^{-}\bar{\nu}_{\mu})$,''
  Phys.\ Rev.\ Lett.\  {\bf 115}, 111803 (2015)
  Addendum: [Phys.\ Rev.\ Lett.\  {\bf 115}, 159901 (2015)]
  doi:10.1103/PhysRevLett.115.159901, 10.1103/PhysRevLett.115.111803
  [arXiv:1506.08614 [hep-ex]].

\bibitem{CCFR}
S.~R.~Mishra {\it et al.} [CCFR Collaboration],
  ``Neutrino tridents and W Z interference,''
  Phys.\ Rev.\ Lett.\  {\bf 66}, 3117 (1991).
  doi:10.1103/PhysRevLett.66.3117

\bibitem{tandean}
  S.~Oh and J.~Tandean,
  ``Rare B Decays with a HyperCP Particle of Spin One,''
  JHEP {\bf 1001}, 022 (2010)
  doi:10.1007/JHEP01(2010)022
  [arXiv:0910.2969 [hep-ph]].

\bibitem{Farzan}
  Y.~Farzan,
  ``A model for large non-standard interactions of neutrinos leading to the LMA-Dark solution,''
  Phys.\ Lett.\ B {\bf 748}, 311 (2015)
  doi:10.1016/j.physletb.2015.07.015
  [arXiv:1505.06906 [hep-ph]].

\bibitem{Aaij:2016flj}
  R.~Aaij {\it et al.} [LHCb Collaboration],
  ``Measurements of the S-wave fraction in $B^{0}\rightarrow K^{+}\pi^{-}\mu^{+}\mu^{-}$ decays and the $B^{0}\rightarrow K^{\ast}(892)^{0}\mu^{+}\mu^{-}$ differential branching fraction,''
  JHEP {\bf 1611}, 047 (2016)
  doi:10.1007/JHEP11(2016)047
  [arXiv:1606.04731 [hep-ex]].

\bibitem{CDFupdate}
\textbf{CDF} Collaboration, \textit{{Updated Branching Ratio Measurements of
  Exclusive $b \to s \mu^+\mu^-$ Decays and Angular Analysis in $B\to
  K^{(*)}\mu^+\mu^-$ Decays }}, . CDF public note 10894.

\bibitem{Chatrchyan:2013cda}
  S.~Chatrchyan {\it et al.} [CMS Collaboration],
  Phys.\ Lett.\ B {\bf 727}, 77 (2013)
  doi:10.1016/j.physletb.2013.10.017
  [arXiv:1308.3409 [hep-ex]].

\bibitem{Khachatryan:2015isa}
\textbf{CMS} Collaboration, V.~Khachatryan et~al., \textit{{Angular analysis of
  the decay $B^0 \to K^{*0} \mu^+ \mu^-$ from pp collisions at $\sqrt s = 8$
  TeV}},  {\em Phys. Lett.} \textbf{B753} (2016) 424--448,
  [\href{https://arxiv.org/abs/1507.08126}{\texttt{arXiv:1507.08126}}].

\bibitem{Aaij:2014pli}
R.~Aaij {\it et al.} [LHCb Collaboration],
  ``Differential branching fractions and isospin asymmetries of $B \to K^{(*)} \mu^+ \mu^-$ decays,''
  JHEP {\bf 1406}, 133 (2014)
  doi:10.1007/JHEP06(2014)133
  [arXiv:1403.8044 [hep-ex]].

\bibitem{Lees:2013nxa}
J.~P.~Lees {\it et al.} [BaBar Collaboration],
  ``Measurement of the $B \to X_s l^+l^-$ branching fraction and search for direct CP violation from a sum of exclusive final states,''
  Phys.\ Rev.\ Lett.\  {\bf 112}, 211802 (2014)
  doi:10.1103/PhysRevLett.112.211802
  [arXiv:1312.5364 [hep-ex]].

\end{thebibliography}
\end{document}